\documentclass{aa}
\def\sqr#1#2{{\vcenter{\vbox{\hrule height.#2pt
    \hbox{\vrule width.#2pt height#1pt \kern#1pt \vrule width.#2pt}
    \hrule height.#2pt}}}}

\def\simlt{\lower.5ex\hbox{$\; \buildrel < \over \sim \;$}}
\def\simgt{\lower.5ex\hbox{$\, \buildrel > \over \sim \,$}}
\def\bull{\vrule height .9ex width .8ex depth -.1ex}

\usepackage{savesym}
\usepackage{amsmath}
\savesymbol{iint}
\usepackage{txfonts}
\restoresymbol{TXF}{iint}
\usepackage{amssymb}
\usepackage{wasysym}
\usepackage{multirow}
\usepackage[dvips]{epsfig}
\usepackage{graphicx}
\usepackage{natbib}
\usepackage{xcolor}
\definecolor{gray}{cmyk}{0.5,0.5,0.5,0}
\begin{document}

   \title{Stellar models simulating the disk-locking mechanism
and the evolutionary history of the Orion Nebula cluster and NGC\,2264%
\thanks{The complete version of Table \ref{tabevol} is only available in electronic form
at the CDS via anonymous ftp to cdsarc.u-strasbg.fr (130.79.128.5)
or via http://cdsweb.u-strasbg.fr/cgi-bin/qcat?J/A+A/}}

%   \subtitle{}

   \author{N.R. Landin\inst{1,2},  
      L.T.S. Mendes\inst{2,3}, L.P.R. Vaz\inst{2} \and S.H.P. Alencar\inst{2} 
          }

   \offprints{N.R.Landin}

   \institute{Universidade Federal de Vi\c cosa - {\it Campus} UFV Florestal, C.P. 35690-000 --
              Florestal, MG, Brazil; \\
              \email{nlandin@ufv.br, nlandin@fisica.ufmg.br}
              \and
              Depto.\ de F\'{\i}sica,
              Universidade Federal de Minas Gerais, C.P.702, 31161-901 --
              Belo Horizonte, MG, Brazil; \\
              \email{lpv@fisica.ufmg.br, silvia@fisica.ufmg.br}
              \and
              Depto.\ de Engenharia Eletr\^onica,
              Universidade Federal de Minas Gerais, C.P.702, 31270-901 --
              Belo Horizonte, MG, Brazil; \\
              \email{luizt@cpdee.ufmg.br}  
             }

   \date{Received \dots; accepted \dots
%Tue Aug  3 16:29:18 BRT 2010
%   }
}

% \abstract{}{}{}{}{} 
% 5 {} token are mandatory

\abstract
{Rotational evolution in young stars is described by pre--main sequence 
evolutionary tracks including non-gray boundary conditions, 
rotation, conservation of angular momentum, and simulations of disk-locking. 
}
{
By assuming that disk-locking is the regulation mechanism for the stellar 
angular velocity during the early stages of pre--main sequence evolution, we
use our rotating models and observational data to constrain disk lifetimes 
($T_{\rm disk}$) of a representative sample of low-mass stars in two young 
clusters, the Orion Nebula cluster (ONC) and NGC\,2264, and to better understand 
their rotational evolution.
}
{
The period distributions of the ONC and NGC\,2264 are known to be bimodal and to
depend on the stellar mass. To follow the rotational evolution of
these two clusters' stars, we generated sets of evolutionary tracks from a fully 
convective configuration with low central temperatures (before D- and 
Li-burning). We assumed that the evolution of fast rotators can be represented 
by models considering conservation of angular momentum during all stages and
of moderate rotators by models considering conservation of angular velocity 
during the first stages of evolution. With these models we estimate a mass 
and an age for all stars.
}
{
The resulting mass distribution for the bulk of the cluster population is 
in the ranges of 0.2-0.4\,$M_{\odot}$ and 0.1-0.6\,$M_{\odot}$ for the ONC and 
NGC\,2264, respectively. For the ONC, we assume that the secondary peak in the 
period distribution is due to high-mass objects still locked in their disks,
with a locking period ($P_{\rm lock}$) of $\sim$8 days. For NGC\,2264 we make 
two hypotheses: (1) the stars in the secondary peak are still locked with 
$P_{\rm lock}$=5 days, and (2) NGC\,2264 is in a later stage in the 
rotational evolution. Hypothesis 2 implies in a disk-locking scenario with 
$P_{\rm lock}$=8 days, a disk lifetime of 1~Myr and, after that, constant 
angular momentum evolution. We then simulated the period distribution of 
NGC\,2264 when the mean age of the cluster was 1~Myr. Dichotomy and bimodality 
appear in the simulated distribution, presenting one peak at 2 days and 
another one at 5-7~days, indicating that the assumption of 
$P_{\rm lock}$=8~days is plausible. Our hypotheses are compared with
observational disk diagnoses available in the literature for the ONC and NGC\,2264, 
such as near-infrared excess, $H\alpha$ emission, and spectral energy distribution 
slope in the mid-infrared.   
}  
{  
Disk-locking models with $P_{\rm lock}$=8 days and 0.2~Myr$\leq$$T_{\rm disk}$$\leq$3~Myr  
are consistent with observed periods of moderate rotators of the ONC.   
For NGC\,2264, the more promising explanation for the observed period 
distribution is an evolution with disk-locking (with $P_{\rm lock}$ near 
8 days) during the first 1~Myr, approximately, but after this, the evolution 
continued with constant angular momentum.   
}  
%{Place here some concluding remarks?\dots}  
  
  
\authorrunning {Landin et al.}  
\titlerunning {Disk-locking and the evolutionary history of ONC and NGC\,2264}  
  
\maketitle  
  
%________________________________________________________________  
  
\section{Introduction}  
  
The evolution of angular momentum of  
pre-main sequence (pMS)  
stars is an important  
problem in star formation and is currently a controversial topic (\citealp{cieza06} and \citealp{meibom13}).  
Gravitational contraction of the stars and only marginal loss of angular momentum  
can explain the spin-up of stars from  
pMS phase to zero-age main-sequence (\textsc{zams})  
and,  then, the existence of rapidly rotating stars on the \textsc{zams}.  
On the other hand, the broad period distribution of \textsc{zams} stars and the existence   
of slow rotators can only be explained if during early phases of evolution  
some stars lost  
significant fractions of their angular momenta, and  
this loss  
is, in  
addition, rather different from star to star \citep{lamm05}.   
  
The observational basis for the study of stellar angular momentum starts  
with the knowledge that at about spectral type F5 a sharp break  
in the rotation  
velocity distributions of field stars occurs. 
\citet{kraft67}  
proposed that rotation declines with advancing age  
as a result of magnetically coupled winds, even after a star reaches the main  
sequence (MS).  
\citet{skumanich72} concluded, based on the available  
observed rotational velocities,  
that  
they  
decrease with time as $t^{-1/2}$.  
\citet{benz84} confirmed that for stars  
later than F8, the rotational velocities decrease with age, but this decay is   
only poorly fitted by the $t^{-1/2}$ law.   
When short-period photometric variations were discovered in young   
clusters, it was found that some stars do not follow the Skumanich law   
(\citealp{alphenaar81} and \citealp{meys82}). As soon as fast rotators were discovered in  
clusters of different ages, an interesting feature became clear: the mass  
dependence of the rapid rotation phenomenon. Several works   
(\citealp{stauffer97b}, \citealp{prosser95}, \citealp{stauffer87a}, \citealp{soderblom93}, \citealp{stauffer87b}, and \citealp{stauffer97a})  
pointed out that rapid rotation disappears  
slower for stars of lower mass.  
The range in rotation for T\,Tauri stars
appeared to be
narrower than for the  
MS, which is problematic if the T\,Tauri star velocity distribution is to  
evolve into that seen on the   
MS.
Since the works by \citet{attridge} and \citet{choi96}, it is well   
known that   
rotational periods of T\,Tauri stars in the Orion Nebula cluster (ONC) show a very  
characteristic distribution. Classical T\,Tauri stars (\textsc{ctts}) have a  
narrow period distribution that peaks at about 8 days. Weak-line  
T\,Tauri  
stars (\textsc{wtts}) have a much broader distribution, with some stars  
showing  
rotational periods lower by a factor of 4 relative to those   
presented by \textsc{ctts}. \citet{herbst02} found that the period distribution for  
ONC stars of spectral type M2\footnote{The corresponding
mass of this spectral type division is model dependent. According to theoretical 
evolutionary tracks by \citet{dantona97}, it corresponds to 0.25M$_{\odot}$
and, according to the models presented in this work, to 0.30M$_{\odot}$.} or earlier   
peaks at 2 and 8 days,  
 while the distribution for ONC stars 
of  
later spectral types 
have a peak at 2 days and only a tail of slow rotators. \citet{lamm05}   
analyzed the period distribution of NGC\,2264 stars as a function of their 
colors (which have a dependence on the masses), but not as a funtion of 
specifically their masses or spectral types. 
They found the same behavior as in  
the ONC: stars redder than a given color (lower mass stars) present only one peak  
in the period distribution, at around 1 day, and stars bluer than a given  
color (more massive stars) have a period histogram with two peaks: one around  
1 day, and another one around 4-5 days.   
  
These observations suggest that T\,Tauri stars in accretion disk  
systems are subjected to a regulation of their angular velocities,  
which counteracts the tendency to spin up  
both from accretion of disk  
material of high specific angular momentum and  
from readjustments in  
the moment of inertia as they contract toward the MS.  
The presence of a circumstellar disk apparently enforces nearly constant  
rotation in the central star, preventing the expected spin-up. The lower rotation  
period of \textsc{wtts} is consistent with an evolution without a circumstellar disk,  
being free to experience the spin-up. On the other hand, the observed range   
of MS rotation rates could be explained  
if a range in  
accretion disk lifetimes is assumed.  
According to \citet{edwards93}, a mechanism proposed by \citet{konigl91}, known as   
``disk-locking'', which consists of a magnetic star-disk interaction, could be   
responsible for this characteristic period distribution in young stellar clusters.   
This is the most acceptable picture of the problem, but \citet{matt04} pointed out  
that observations by \citet{stassun99} and \citet{johnskrull99} and theoretical considerations  
of \citet{safier98} have called the standard disk-locking scenario into question.   
Based on the work by \citet{uzdensky02}, which demonstrated that differential rotation  
between the star and disk leads to an opening of the field lines that drastically  
reduces the magnetic connection between the two, \citet{matt04} emphasized that  
the disk-locking model cannot account for the  
angular momentum loss of the slow
rotators, concluding
that accretion-powered stellar winds are a promising scenario for solving the stellar angular 
momentum problem. On the other hand, \citet{cauley12} derived accretion parameters for 36 
T\,Tauri stars in NGC\, 2264 and analyzed different models of magnetospheric accretion and
found good support for disk-locking, but not for theories that assume a dipolar magnetic field 
geometry. 
In addition,
their
results found no support for accretion-powered
stellar winds, although
they
agree that this mechanism certainly provides a mean of angular momentum
removal for \textsc{ctts}.

Several works were carried on to examine the hypothesis that  
disk-locking regulates
pMS angular momenta. These works were 
based on at least one observational property that indicates the presence 
of a circumstellar disk. \citet{rebull00} described the origin of some disk 
indicators commonly used in the literature, such as U-V excess, 
$H\alpha$ emission (which are accretion diagnostics and, 
consequently, indicate the presence of circumstellar disks), and infrared 
excesses ($I$$-$$K$ and $H$$-$$K$, which are diagnostics of
dust heated in circumstellar disk). \citet{rebull02}
used $U$$-$$V$, $I$$-$$K$, $H$$-$$K$ 
excesses, and $H\alpha$ emission to identify circumstellar disk candidates 
in NGC 2264.
\citet{makidon04} used these same disk indicators to 
search for 
eventual correlations with period 
of NGC 2246 stars, but
found no conclusive evidence that
slower rotating 
stars have disk indicators, or that faster
rotators are less likely to have
disk indicators. \citet{dahm05} analyzed the NGC\,2264 rotational data from 
\citet{lamm04} and \citet{makidon04}
and found a weak-to-moderate positive 
correlation between $H$$-$$K$ color and rotation periods for the \textsc{ctts}, in the
sense that
longer period objects tend to have larger $H$$-$$K$ colors.
They found a similar positive correlation between 
$L_{\rm H\alpha}$, $H\alpha$ luminosity, and
rotation period among the \textsc{ctts}. By using 
near-infrared excesses to identify disk candidates in the ONC, \citet{hillen98} 
concluded that this disk indicator misses about 30\% of disks that can be 
detected at longer wavelengths. Then, in the attempt to use a more reliable 
disk indicator,
\citet{cieza07} used data from the 
{\it Spitzer} Space Telescope in the IRAC band passes to test the star-disk interaction paradigm in
NGC\,2264 and the ONC.
By using {\it Spitzer} mid-infrared data, they
observed
a clear increase
in the disk fraction with period in both clusters across the entire period range
populated by
their members. They also showed that the peak at 8 days of the  
distribution of ONC stars of spectral type M2 or earlier is dominated by stars possessing a disk,  
while the peak at 2 days is dominated by stars without  
disks. Their results   
present strong evidence that star-disk interaction regulates the angular   
momentum of these young stars. On the other hand, \citet{cieza06} made the   
same analysis for IC\,348, a 1-3\,Myr old cluster, and found no evidence that the tail of slow 
rotators
(stars with $M$$<$0.25M$_{\odot}$)
or the long-period peak
($M$$>$0.25M$_{\odot}$)
are preferentially  
populated by objects with disks,  
as might be expected based on the current  
disk-braking model. In addition, they found no significant correlation between period  
and the  
IR-excess, regardless of the mass range considered. 
These results are in conflict with those by \citet{nordhagen06},  
who found that  
the variability observed in IC 348 \textsc{wtts} arises primarily from the rotation of a spotted  
photosphere and that the irregular variability of \textsc{ctts} is due to the presence  
of an accretion disk. \citet{nordhagen06} have also shown that the  
solar mass range stars  
present in IC\,348 have a rotation period distribution  
similar to that of the ONC,  
but contain a  
larger portion of slow rotators than  
is seen in NGC\,2264. As \citet{nordhagen06} considered an age 
of 3~Myr for IC\,348, for them the IC\,348 age is closer to NGC\,2264 than to the 
ONC. Then, they concluded that another factor in addition to age and mass is probably 
important in establishing the rotation properties within a cluster, and their 
suggestion  is that this factor is the environment. 
However, \citet{affer13} suggested that the results found 
by \citet{cieza06} suffer from several biases affecting the selected samples 
and even the disk and accretion indicators adopted.  
Based on CoRot light curves, \citet{affer13} determined
high-accuracy rotation periods for 189 stars of NGC\,2264. By using the $H\alpha$
equivalent width, they found that rotational distributions of \textsc{ctts} 
and \textsc{wtts} are different, consistent with the disk-locking scenario.
 
To learn more about the initial conditions, it is 
ne\-cessary to obtain observational data for pMS stars. Through a monitoring
campaign of some open clusters in star-forming regions, ``The Monitor Project'' researchers
are trying to further our understanding of the angular momentum of young, 
low-mass stars \citep{aigrain07}. This survey
measured photometric 
rotation periods for a large number of objects at a range of ages from
the stellar birth line to the zero-age main sequence belonging to some young clusters 
\citep{irwin07a,irwin07b,irwin08a,irwin08b}. 

From the theoretical point of view,  
many
of the angular momentum evolution  
models
were developed in an attempt to address these questions.  
\citet{kawaler87} used an extrapolation of the \citet{kraft70} relation to infer  
the initial  
pMS angular momentum as a function of mass. \citet{pinson89,pinson90} constructed  
solar models with rotation  
using the   
\citet{kawaler87} relations as initial conditions, and \citet{kawaler88}   
presented a formulation for  
the angular momentum loss by magnetic stellar winds.  
Microscopic diffusion effects were  
considered in the models  
by \citet{chaboyer95a,chaboyer95b}.  
Several authors explored the rotational evolution of solar analogs under a variety of   
assumptions about  
both the timescale for rotational coupling between the core  
and envelope,  
and  
the initial rotation rates  
(e.g., \citealp{macgregor91} and \citealp{keppens95}).
The role played by  
star-disk interaction in the origin of the initial conditions   
for stellar rotation was explored by
	\citet{jianke93} and \citet{cameron95}. 
\citet{bouvier97} presented models  
for surface rotation of solar-like stars  
considering nearly solid-body rotation,  
pMS disk-locking, and wind  
braking, using  
the stellar moments of inertia and  
radii from the non-rotating stellar evolution  
model of \citet{forestini94}. They concluded that a 
median disk lifetime of 3~Myr would be necessary to make slow rotators, but 
long disk-coupling times ($\sim$10-20~Myr) are needed 
to explain the rotation rates of the slowest rotator 
stars. By using the Yale stellar evolution 
code\footnote{Several authors cited in this work used the Yale Rotating 
Stellar Evolution Code (YREC) to study angular momentum evolution, for 
instance, \citet{pinson89,pinson90}, \citet{guenther92},  
\citet{chaboyer95a}, \citet{krishnamurthi97}, \citet{barnes99}, and 
\citet{barnes01}.}, \citet{guenther92} and \citet{krishnamurthi97} constructed angular momentum  
models and examined the effects of different assumptions about initial 
conditions on the angular momentum evolution of sotlar-type stars. For 
rotating models, they used  
moments of inertia calculated by non-rotating models.  
For differentially rotating models, modest disk lifetimes
on the order of 3~Myr were needed to produce slow rotators, and for solid-body 
models, longer disk lifetimes (on the order of 20~Myr) were needed, in agreement with
\citet{bouvier97}.
\citet{barnes99} and \citet{barnes01}, also using the Yale code, constructed rotating models  
accounting  
for transport and redistribution of  
both chemical species and  
angular momentum within a star,  
as a result of various rotationally induced instabilities.  
To mimic the effect of disk-locking on the  
pMS, their models were allowed to  
lock the rotation rate of a young star on the Hayashi track for a specified  
time. Their models  
begin at the deuterium MS.   
\citet{barnes99} interpreted rotation period and $B$$-$$V$ color
data of young stars (30~Myr), present in the young clusters IC\,2602
and IC\,2391, in terms of these models. 
The ultrafast rotators (periods as slow as $\sim$0.2~days) could be 
reproduced by assuming an initial rotation period of 4~days and no 
locking phase.  
For the majority of solar-type stars considered in their analysis, they 
explained the rotation period distribution with an initial rotation period 
of 16~days and disk lifetimes of up to 5~Myr.
  
Despite all these efforts, no model to date has been able to explain the presence of fast and slowly rotating   
stars in the same young cluster by a single mechanism intrinsic to the star.  
The introduction of rotation in the  
\textsc{aton\,2.3}  
stellar evolutionary code  
makes it possible to explore the physical mechanism that apparently maintains  
the stellar angular velocity constant.  
  
We present our new rotating evolutionary tracks,  
including a simulation of the disk-locking mechanism, as a plausible  
explanation for the rotational evolution of slow rotator  
pMS stars.  
We do not  
model the star-disk interaction,  
but  
only  
mimic its  
effect  
on the stellar rotation by locking  
it for a while.   
The advantage of our models compared to those of   
\citet{bouvier97} and \citet{krishnamurthi97} is that we model not only  
the angular momentum, but also the structure and evolution of  
stars, and our moments of inertia are obtained by taking into account the rotation  
effects. This was also the approach followed by \citet{barnes01}, but they   
began the calculations at the deuterium main sequence, disregarding the  
action of the thermostat of deuterium burning, which may
introduce
numerical uncertainties in stellar ages (see \citealp{landin06} for a discussion).   
Our models instead reach the deuterium-burning phase by contracting in   
thermal equilibrium.   
  
We consider a range of disk lifetimes and of initial   
rotation rates, and use our theoretical predictions  
to analyze the rotational properties of the  
ONC   
and NGC\,2264 stars, which were found to vary considerably with mass.  
Objects with masses higher than a threshold value have a clear bimodal period  
distribution (which peaks $\sim$2 and 8 days for the ONC  
and   
$\sim$1 and 4-5 days for NGC\,2264),  
while the less  
massive sample of each cluster
only constitutes a tail of slow rotators.   
Assuming that  
disk-locking is responsible for the peak at 8 (4-5) days for the ONC (NGC\,2264),   
we investigate the role of disk lifetimes on the rotational evolution of the   
stars in these two young clusters.  
  
In Sect.~\ref{models} we present and describe our disk-locking models  
that we used to follow the evolution  
of moderate rotators present in  
the ONC and NGC\,2264 clusters.   
Models with conservation of angular  
momentum during all evolution (also used in this work) are also
described in   
Sect.~\ref{models}.  
The observational evidence of an angular  
momentum regulation mechanism in  
pMS  
is discussed in  
Sect.~\ref{discussion}.  
The ONC and NGC\,2264 data  
used, the derivation of masses and ages,  
the  
results  
of the rotational period distributions (of both clusters), and  
a discussion of  
the possible interpretations  
are presented in Sect.~\ref{results}. Our conclusions are given in Sect.~\ref{conclusions}.

\section{Models}\label{models}  
  
In this section we briefly present the version of the  
\textsc{aton} code we used  
and show how we introduced the effect of the presence of a disk in our models.   
The angular momentum evolution models, which are generally used in the literature  
to study the rotational evolution of low-mass stars, use as input parameters the   
moments of inertia and radii computed by stellar evolutionary models   
without rotation. The angular momentum evolution is calculated separately
in another code (for example, see \citealp{bouvier97} and \citealp{krishnamurthi97}).   
We used evolutionary models instead that include   
rotation. This is a more consistent approach
because the structural  
effects caused by rotation are, thereby, taken into account. Some structural effects of rotation are discussed in this section.  
The evolutionary tracks  
were computed in the mass range of 0.15-0.8\,M$_{\odot}$.  
The solar chemical composition was adopted  
(Y=0.27 and Z=0.0175). We used the same boundary conditions as \citet{allard1} and   
\citet{allard00}. Convection was treated  
through the classical  
mixing length theory (\textsc{mlt}), with $\alpha$=2.0.  
Since most of the ONC and NGC2264  
objects are   
low-mass  
pMS stars,  
rigid body rotation was assumed.  
We used the opacities reported by \citet{iglesias93} and \citet{alexander94} and the equations of state from \citet{rogers96}  
and \citet{mihalas88}. The  
tracks start from a fully  
convective configuration with central temperatures in the range  
5.3$<$$\log T_\mathrm{c}$$<$5.8,  
follow deuterium and lithium burning, and  
end at the MS configuration.   
For more details see \citet{landin06}.  
  
\citet{attridge} and \citet{choi96} suggested that the mechanism acting on stars  
against the increase in their angular velocity during the  
pMS  
contraction could be disk-locking, due to magnetic coupling between  
the star and the disk \citep{konigl91}.  
This version of the  
\textsc{aton} code  
allows the evolution of stars with disk-locking simulation
by   
keeping their angular velocity constant during the early stages  
of  
pMS.   
During this  
phase, the period of the  
star is fixed at a given   
locking-period, $P_\mathrm{lock}$.  
After this locked phase, the star  
is free to spin up, following a constant angular momentum evolution.   
For ONC stars, we  
used $P_\mathrm{lock}$=8\,days,  
following the suggestion by   
\citet{herbst02}. For NGC\,2264,  
we used $P_\mathrm{lock}$=5\,days,   
based on the value of the second peak in the rotation period  
distribution of this cluster \citep{lamm05}.   
These two
values of
$P_{\rm lock}$ have corresponding values of initial  
angular momentum, which are used as input parameters by the model.  
Based on these initial rotation rates, we  
generated   
sets of evolutionary tracks taking into account the disk-locking   
mechanism, with disk lifetimes of 0.2, 0.5, 1.0, 3.0, and 10.0 Myr.   
 To  
make a comparison with the evolution of stars without accretion disks,  
we  
generated another set of models considering   
angular momentum conservation since the beginning. In these cases,  
the initial angular momentum is a function of mass and is given   
by \citep{kawaler87}  
\begin{equation}  
\centering  
J_\mathrm{kaw}=1.566 \times 10^{50} \left( {M \over M_{\odot}} \right) ^{0.985}  
~~~\mathrm{cgs}.  
\label{kaweq}  
\end{equation}  

It is important to emphasize that the results obtained with  
our rotating models depend on the initial angular momentum (or angular 
velocity) value, which is an input parameter of our models. 
For models with conservation of angular momentum from the  
beginning we used the \citet{kawaler87} relations (Eq. \ref{kaweq}),   
and for disk-locking models we used an initial angular momentum that    
corresponds to a period of 5 or 8 days.   

In Table~\ref{tabevol} we present the 0.4\,M$_{\odot}$ models as   
an example of our calculations with the   
\textsc{aton} code version that includes disk-locking.  
  
\begin{table}[t]  
\footnotesize  
\caption{Pre-main sequence evolutionary tracks  
for   
0.4\,M$_{\odot}$ star for disk-locking models with P$_{\rm lock}$=5~days and T$_{\rm disk}$=1\,Myr$^a$.  
Column 1 gives the logarithm of stellar   
age  
(years); Col. 2 the logarithm of stellar luminosity  
(solar   
units); Col. 3 the logarithm of effective temperature  
(K);  
Col. 4 the mass contained in the convective envelope relative to the   
total mass ($M_\mathrm{conv}/M_\mathrm{tot}$); Col. 5 the logarithm of   
total rotational inertia ($\log I_{\rm tot}$, cgs); Col. 6 the logarithm of rotational inertia 
in the convective envelope ($\log I_{\rm conv}$, cgs);   
and Col. 7 the rotational period ($P_\mathrm{rot}$, days).  
}  
\label{tabevol}  
\centering  
\begin{tabular}{rrccccc}  
\noalign{\hrule\vskip 1pt\hrule\smallskip}  
${\log Age\atop{\mathrm{(years)}}}$ & $\log\frac{L}{L_{\odot}}~~$ & ${\log T_{\mathrm{eff}}\atop{\mathrm{(K)}}}$ & $\frac{M_\mathrm{conv}}{M_\mathrm{tot}}$ & ${\log I_\mathrm{tot}\atop{\mathrm{(cgs)}}}$ & ${\log I_\mathrm{conv}\atop{\mathrm{(cgs)}}}$ & ${P_\mathrm{rot}\atop{\mathrm{(days)}}}~~$ \\  
\noalign{\smallskip\hrule\vskip 1pt}  
 2.6990 &  1.1983 & 3.5364 & 1.000 & 55.8939 & 55.8939 & 5.000 \\ [-1.5pt]  
 3.7669 &  0.8332 & 3.5731 & 1.000 & 55.4176 & 55.4176 & 5.000 \\ [-1.5pt]  
 4.3476 &  0.6145 & 3.5792 & 1.000 & 55.1964 & 55.1964 & 5.000 \\ [-1.5pt]  
 4.7922 &  0.3949 & 3.5801 & 1.000 & 54.9815 & 54.9815 & 5.000 \\ [-1.5pt]  
 5.6514 &  0.1908 & 3.5778 & 1.000 & 54.7917 & 54.7917 & 5.000 \\ [-1.5pt]  
 5.8159 &  0.0243 & 3.5726 & 1.000 & 54.6489 & 54.6489 & 5.000 \\ [-1.5pt]  
 5.9375 & -0.1792 & 3.5615 & 1.000 & 54.4934 & 54.4934 & 5.000 \\ [-1.5pt]  
 6.1166 & -0.3993 & 3.5479 & 1.000 & 54.3315 & 54.3315 & 3.939 \\ [-1.5pt]  
 6.3691 & -0.6192 & 3.5404 & 1.000 & 54.1459 & 54.1459 & 2.569 \\ [-1.5pt]  
 6.6674 & -0.8394 & 3.5367 & 1.000 & 53.9446 & 53.9446 & 1.616 \\ [-1.5pt]  
 6.9887 & -1.0597 & 3.5353 & 1.000 & 53.7342 & 53.7342 & 0.996 \\ [-1.5pt]  
 7.3155 & -1.2801 & 3.5342 & 0.968 & 53.5215 & 53.5208 & 0.610 \\ [-1.5pt]  
 7.6562 & -1.4993 & 3.5341 & 0.758 & 53.3016 & 53.2757 & 0.368 \\ [-1.5pt]  
 8.0368 & -1.6855 & 3.5380 & 0.486 & 53.0810 & 52.9713 & 0.221 \\ [-1.5pt]  
10.0536 & -1.7186 & 3.5372 & 0.520 & 53.0592 & 52.9586 & 0.210 \\ [-1.5pt]  
10.5776 & -1.6919 & 3.5388 & 0.474 & 53.0648 & 52.9477 & 0.213 \\ [-1.5pt]  
10.8272 & -1.6613 & 3.5410 & 0.427 & 53.0680 & 52.9327 & 0.215 \\ [-1.5pt]  
10.9838 & -1.6247 & 3.5438 & 0.384 & 53.0713 & 52.9176 & 0.216 \\ [-1.5pt]  
11.0939 & -1.5809 & 3.5474 & 0.346 & 53.0757 & 52.9065 & 0.219 \\ [-1.5pt]  
11.1772 & -1.5275 & 3.5518 & 0.312 & 53.0822 & 52.8996 & 0.222 \\ \hline  
\multicolumn{7}{p{0.95\columnwidth}}{$^a$ \scriptsize  
The complete version of the table,   
including 40 tracks for 0.15, 0.2, 0.3, 0.4, 0.5, 0.6, 0.7, and 0.8\,M$_{\odot}$  
for models considering  
both  
conservation of angular   
momentum and disk-locking  
(with locking period of 5 and 8~days and disk lifetimes of 0.5 and 1\,Myr) is available only in electronic form.}  
\end{tabular}  
\end{table}  

\begin{figure}[t]  
\centering{  
\includegraphics[width=9cm]{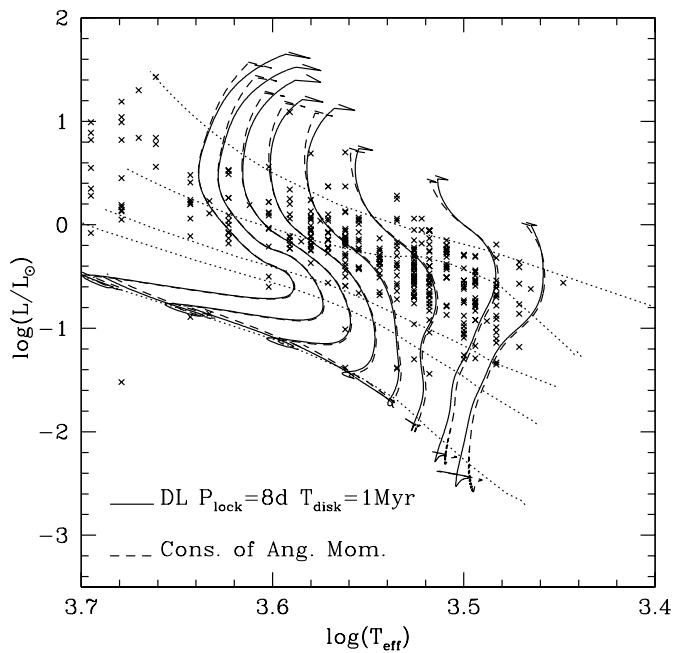}  
}  
\caption{Evolutionary tracks for  
0.15, 0.2, 0.3, 0.4, 0.5, 0.6, 0.7, and 0.8\,M$_{\odot}$ (from low  
to high temperatures), considering constant angular  
momentum evolution  
(dashed lines) and  
disk-locking evolution with P$_{\rm lock}$=8~days and T$_{\rm disk}$=1\,Myr  
(solid lines). Dotted lines   
represent the isochrones  
for 0.1, 1, 5, 10, and 100\,Myr (from high to low  
luminosities). Crosses represent  
the ONC stars. Effective temperatures and luminosities were
taken from \citet{hillen97}.  
}  
\label{tracks}  
\end{figure}  
  
\begin{figure}[t]  
\centering{  
\includegraphics[width=8.5cm]{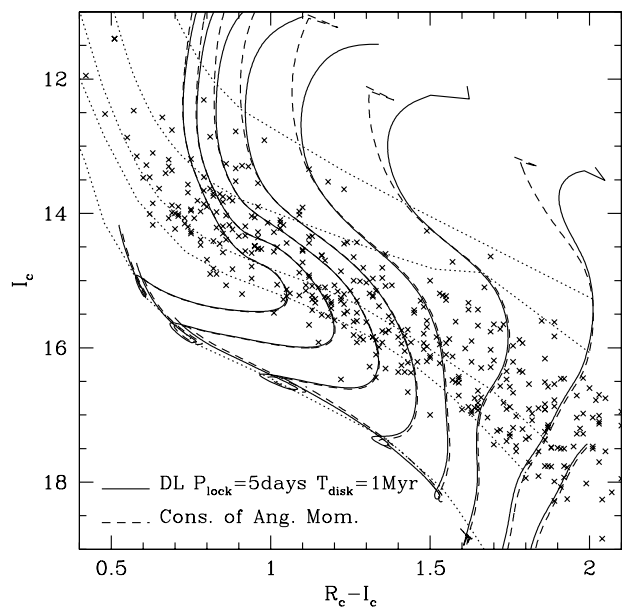}  
}  
\caption{Color-magnitude  
diagram ($I_c$ vs.\ $R_c$$-$$I_c$) and our evolutionary tracks for models  
of 0.15, 0.2, 0.3, 0.4, 0.5, 0.6, 0.7, and 0.8\,M$_{\odot}$ (from right  
to left)  
considering constant angular momentum  
evolution (dashed lines) and evolution with disk-locking with P$_{\rm lock}$=5~days and T$_{\rm disk}$=1\,Myr (solid lines).   
Isochrones  
(dotted lines) for 0.1, 1, 5, 10, and 100\,Myr  
are shown (from top to bottom). Crosses are NGC\,2264 stars \citep{lamm04}.  
}  
\label{ngctracks}  
\end{figure}

\begin{figure}[htb]  
\centering{  
\includegraphics[width=9.3cm]{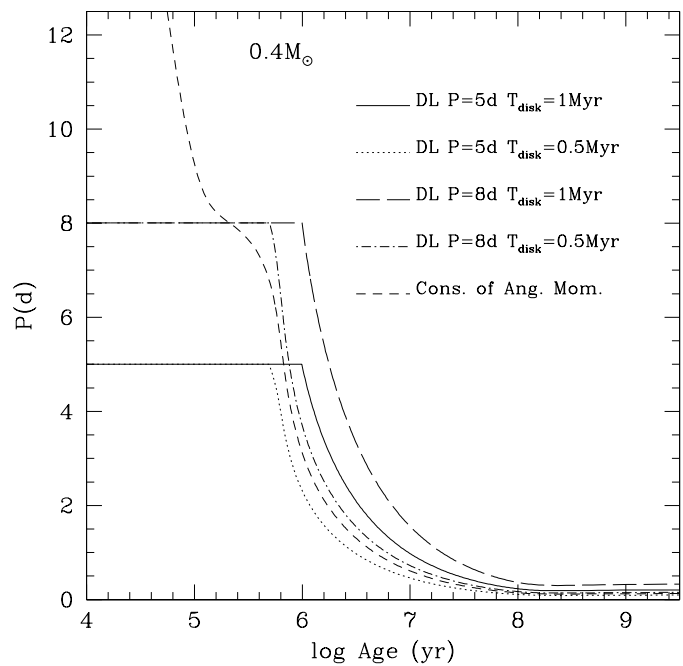}  
}  
\caption{Time evolution of the rotation periods for some 0.4$\rm M_{\odot}$ stellar models.  
}  
\label{perevolution}  
\end{figure}  
  
In Figs.~\ref{tracks} and \ref{ngctracks} we show   
the evolutionary tracks obtained with different sets of models.  
We  
note that a  
difference in the  
rotation evolution mainly affects low-mass tracks. Even a short disk lifetime   
of 1\,Myr can shift the evolutionary paths followed by the tracks. In the beginning  
of the Hayashi phase (higher luminosities), the evolutionary tracks obtained with disk-locking models  
are slightly cooler than those  
obtained with conservation of angular momentum.  
When the tracks approach the MS (lower luminosities), 
the situation is reversed (Figs.~\ref{tracks}, \ref{ngctracks}). This behavior can be connected  
with a structural effect caused by rotation, the mass-lowering effect, which produces  
a shift in the evolutionary tracks of rotating stars toward lower effective  
temperature and luminosity, simulating a non-rotating star of lower mass \citep{mendes99}.  
This effect is more or less evident, depending on the intensity of rotation.  
In
Fig.~\ref{perevolution}
we show as an example the time evolution of   
rotation periods of 0.4\,$\rm{M_{\odot}}$ models.   
Rotation periods as a function of age are  
shown for models conserving angular momentum since the beginning and for disk-locking (DL) 
models with $P_{\rm lock}$=5\,days and $T_{\rm lock}$=0.5\,Myr,   
$P_{\rm lock}$=5\,days   
and $T_{\rm lock}$=1\,Myr, $P_{\rm lock}$=8\,days and $T_{\rm lock}$=0.5\,Myr, and   
$P_{\rm lock}$=8\,days and $T_{\rm lock}$=1\,Myr.  
In this work, models with conservation of angular momentum start the evolution with angular
velocities given by Kawaler's relation (Eq. \ref{kaweq}), which 
are lower than the locking velocities used in our 
disk-locking models for a given mass. This behavior is shown in Fig. \ref{perevolution}, which plots the temporal evolution of periods, instead of velocities. For this reason, in the first stages   
of evolution (ages$<$1 Myr), the  
tracks produced by the models conserving angular momentum   
seem to be
more massive non-rotators
than the   
disk-locking models, and then seem to have   
higher effective temperatures.  
For ages greater than $\sim$1~Myr, the tracks of models with conservation of angular momentum   
evolve to configurations whose angular velocity is higher than disk-locking models with  
$T_{\rm disk}$=1~Myr and $P_{\rm lock}$=5~days, $T_{\rm disk}$=1~Myr and   
$P_{\rm lock}$=8~days, and $T_{\rm disk}$=0.5~Myr and $P_{\rm lock}$=8~days. Consequently,  
the tracks of models conserving angular momentum exhibit lower effective temperatures than the  
disk-locking
models for the same mass, except for DL models 
with $T_{\rm disk}$=0.5~Myr and $P_{\rm lock}$=5~days.
  
\begin{figure}[t]  
\centering{  
\includegraphics[width=9.3cm]{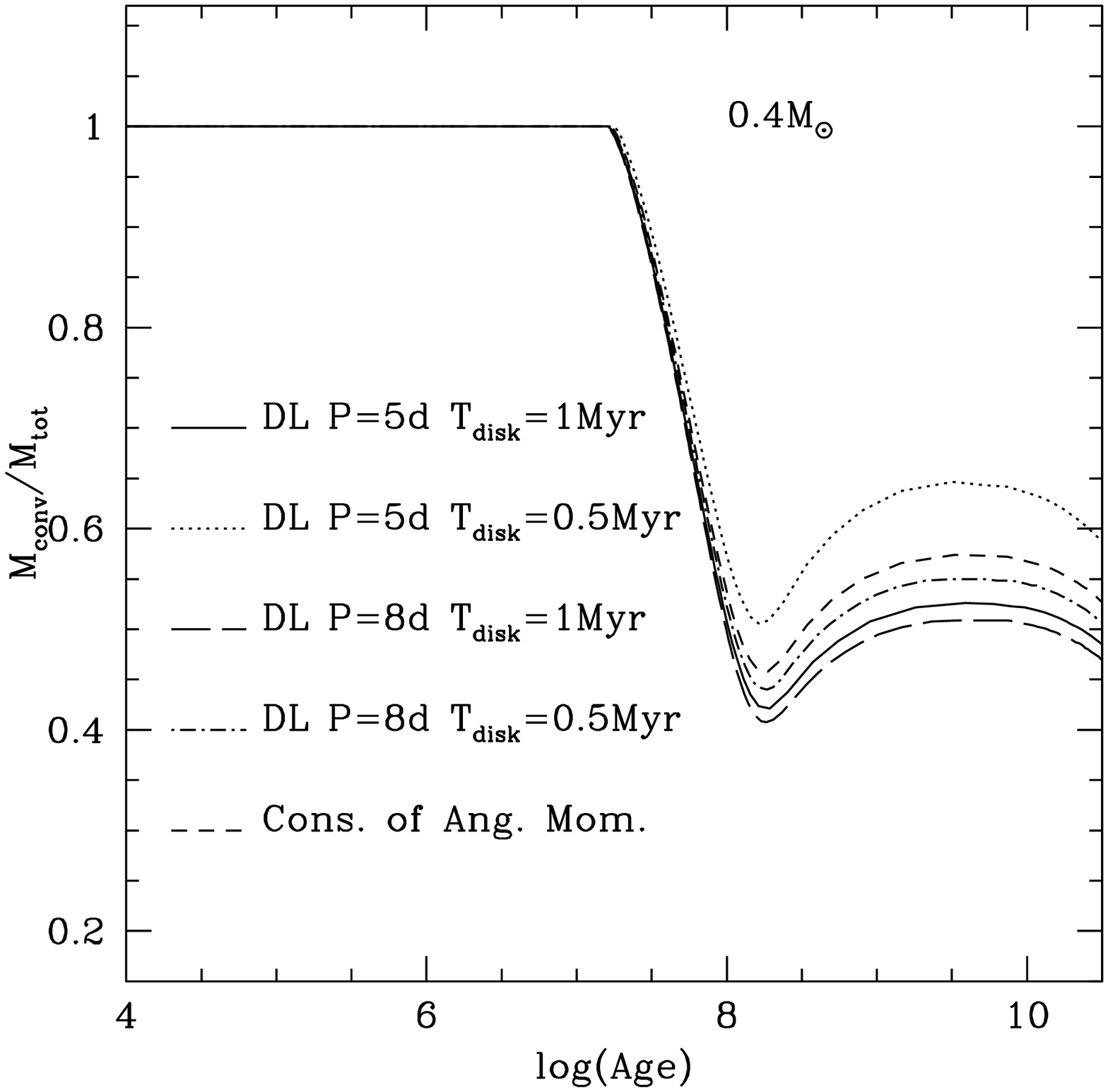}  
}  
\caption{Time evolution of   
the convective envelope relative mass  
for a 0.4M$_{\odot}$ model   
considering constant angular momentum evolution  
and evolution with disk-locking for different disk lifetimes.}  
\label{mconv}  
\end{figure}  
  
The impact of rotational evolution on the stellar structure   
can also be seen in Fig.~\ref{mconv}, where we show the mass of the convective envelope of a 0.4M$_{\odot}$ model as a function of  
stellar age.   
During the first Myr, 0.4$\rm M_{\odot}$ stars are completely convective, and so   
the relative mass of the convective envelope ($\rm{M_{conv}/M_{tot}}$) is equal   
to 1. From this age on, the mass contents of the convective envelope changes   
depending on the kind of rotation evolution followed by the star.     
By comparing rotation periods and the mass of the convective envelope at   
a given age (Figs. \ref{perevolution} and \ref{mconv}),  
we note that the shorter the rotation period (or the greater the angular  
velocity), the larger the mass contents of the convective envelope.   
This behavior can also be understood by recalling the mass-lowering effect.
The faster the star rotates, the more it mimics a lower mass star, and the 
larger its envelope.
We discussed the case of 0.4$\rm M_{\odot}$ models, but this is also valid for other  
mass models,
with small
differences. The relative mass of convective 
envelopes, for instance, is less sensitive to the rotational evolution model as the stellar mass increases.  
  
\begin{figure}[t]  
\centering{  
\includegraphics[width=9cm]{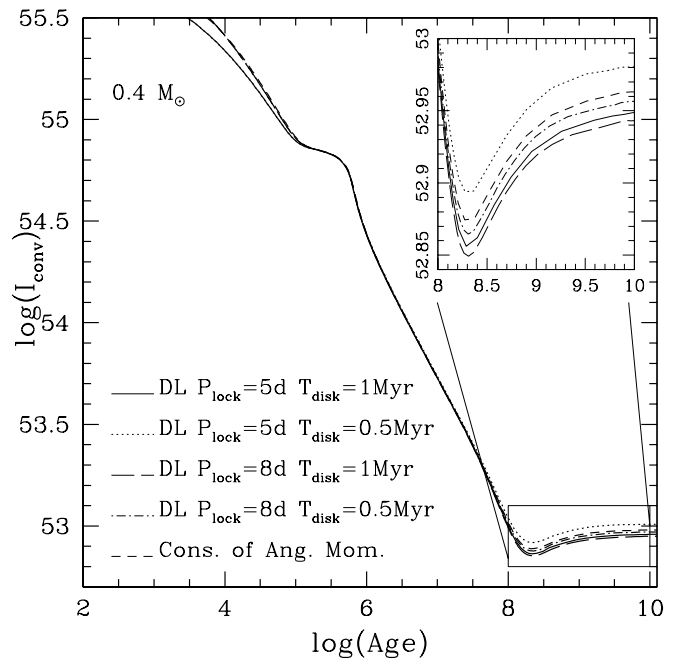}  
}  
\caption{Rotational inertia in the convective envelope versus stellar age  
of 0.4\,M$_{\odot}$ models considering constant angular 
momentum evolution and evolution  
with  
disk-locking for different disk lifetimes. 
}  
\label{irotage}  
\end{figure}

Young stars are nearly completely convective and have large moments of inertia  
(or rotational inertia). As they contract, approaching the main-sequence,  
their surface convection zones retreat and their total rotational 
inertia, $I_{tot}$, decreases. A fraction of $I_{tot}$, the 
rotational inertia in the convection zone of the stars, $I_{conv}$, also decreases. 
This general behavior is visible Fig.~\ref{irotage}, which shows  
the temporal evolution of rotational inertia in the convective envelope 
for 0.4\,M$_{\odot}$ stars. During the very early stages of evolution (ages
shorter than 0.1~Myr), models with conservation of   
angular momentum since the beginning produce higher rotational   
inertia in the convective envelope, followed by disk-locking models with P$_{\rm lock}$=8~days and  
disk-locking models with P$_{\rm lock}$=5~days, in this order. A detailed   
inspection in the calculations reveals that values of rotational inertia 
in the convective envelope    
calculated with disk-locking models with the same locking period do not vary  
significantly. However, for more advanced stages of evolution (ages greater than  
100~Myr) higher rotational inertia in the convection zone is produced by DL models with   
P$_{\rm lock}$=5~days and T$_{\rm disk}$=0.5~Myr, followed by models   
conserving angular momentum since the beginning, DL models with   
P$_{\rm lock}$=8~days and T$_{\rm disk}$=0.5~Myr, DL models with  
P$_{\rm lock}$=5~days and T$_{\rm disk}$=1~Myr, and DL models with  
P$_{\rm lock}$=8~days and T$_{\rm disk}$=1~Myr. This is shown
in the upper right   
corner of Fig.~\ref{irotage}, where we display an amplified view of  
rotational inertia in the convection zone obtained by the models for ages greater than 100~Myr.  
This behavior of $I_{conv}$ for different 
rotational evolutions is also valid for other mass values,  
with small differences mainly for 0.8\,M$_{\odot}$.    
  
\section{Discussions}\label{discussion}  
  
The ONC and NGC\,2264 are the two best studied young star clusters, including analyses   
at different wavelengths. For  
the ONC we can cite, for example,  
\citet{hillen97}, in the optical; \citet{hillen98}, in the near-infrared;   
\citet{cieza07} in the mid-infrared;
\citet{flacco03a,flacco03b} and \citet{stassun04}, in the X-ray;  
\citet{stassun99} and \citet{herbst02} in photometric variability studies to determine   
rotational periods.  
For NGC\,2264, \citet{rebull02} conducted studies in the optical and   
near-infrared; \citet{cieza07} and \citet{dahm12} analyzed it in the 
mid-infrared; \citet{flacco99,flaccomio06} and \citet{rebull06} investigated the  
cluster in X-rays; \citet{teixeira12} analyzed the nature of the disk population of NGC\,2264 stars by using {\it Spitzer} data; \citet{venuti14} characterized the accretion process of NGC\,2264 stars by performing deep {\it ugri} mapping and {\it u}- and {\it r}-band monitoring with CFHT/MegaCam; \citet{lamm04}, \citet{makidon04} 
and \citet{affer13} performed photometric  
variability studies and identified several periodic and irregular variables. 
\citet{favata10} and \citet{affer13} observed NGC\,2264 with the CoRoT space 
telescope, which allowed them to determine rotation periods ($P$$<$12\,days) with
high accuracy and free of the biases imposed by ground-based observations.
Preliminar results from \citet{favata10} showed that stars with M$>$0.25M$_{\odot}$ have a 
single-peaked distribution and no short-period peak. The short-period 
peak ($\sim$1 day) present in the ground-based distribution appeared to be 
spurious. However, \citet{affer13} compared rotation periods obtained with
CoRoT and ground-based data of \citet{lamm04} for 103 stars common in both
samples. They found that 83\% of the sample have consistent rotation periods,
with the exception of only five stars that had one-day periods erroneously determined by the ground-based 
method.

This clearly shows that
similar characteristics were detected in the ONC and NGC\,2264, and 
some works are dedicated  
to compare their rotational properties   
(\citealp{lamm05} and \citealp{cieza07}). This is also one of the purposes of  
the present work.  
  
\textsc{ctts} were found to be remarkably slow rotators,  
having lost large   
amounts of angular momentum during the protostar or early  
pMS phase  
(\citealp{bouvier86} and \citealp{hartmann86}),  
and the magnetic coupling between star and  
disk was proposed as the dominant mechanism of this process.   
Observational support for this picture includes  
evidence for a correlation of rotation properties with the presence of disks  
(\citealp{bouvier93} and \citealp{edwards93}).  
This disk-locking scenario was observationally supported by rotation  
period studies in the ONC, where \citet{attridge} first discovered a  
bimodal period distribution (with peaks around 2 and 8 days) and 
\citet{herbst01,herbst02}, in addition to confirming this, emphasized that it occurs for stars
of spectral types earlier than M2.  
The bimodality observed in such stars 
was interpreted as an effect of   
disk-locking with a locking period of about 8 days, while stars with shorter  
periods are presumably not locked to their disks. \citet{stassun99} were unable to confirm the existence of a bimodal period distribution in the ONC and   
did not find differences between the rotation periods for \textsc{ctts} and \textsc{wtts}.  
This probably is a consequence of their study  
being strongly biased toward   
periods shorter than or equal to 8 days and toward lower mass stars. In   
addition, they did not analyze the influence of mass in the period 
distribution. 
  
\citet{makidon04} and \citet{lamm05} studied the period distribution of  
pMS stars  
in NGC\,2264. While the first authors have not observed bimodality in the  
total period distribution, Lamm and collaborators found it  
to be highly color dependent. They also reported that the period distribution is bimodal for bluer stars ($R_c$$-$$I_c$$<$$1.3$, which corresponds to M$>$0.25M$_{\odot}$ according to \citealt{dantona97} models) and unimodal for redder  
ones ($R_c$$-$$I_c$$>$$1.3$ or M$<$0.25M$_{\odot}$).  
\citet{lamm05}  
noted that the median rotation periods of the bluer and redder     
stars in the ONC, a younger cluster,  
are 6.75  
and 3.3 days, respectively, which means that stars with M$<$$0.25M_{\odot}$    
rotate on average faster by a factor of 2 than those with M$>$$0.25M_{\odot}$.  
However, for NGC\,2264,  
an older cluster than the ONC, they found that the median rotation periods of    
the bluer and redder stars are 4.7  
and 1.9 days, respectively, which means that on average, stars with M$<$$0.25M_{\odot}$   
rotate faster than those with M$>$$0.25M_{\odot}$ by a factor of 2.5.  
They estimated an age ratio of about 2 between the two clusters on the basis of  
pMS  
models. According to our models (with and without disk-locking),  
the estimate for this ratio is around 3. In addition,  
\citet{lamm05} found that the period distribution of NGC\,2264 is similar in  
form to that of the ONC, but shifted to shorter periods. Their quantitative comparisons  
between the period distributions of the two clusters suggest that about 80\%  
of the stars have spun up from the age of the ONC to the age of NGC\,2264.   
Based on this  
suggestion  
and  
on the estimated age ratio between the two clusters,  
they found that the average spin-up by a factor of 1.5-1.8 from the age of the ONC  
to the age of NGC\,2264 is consistent with a decreasing stellar radius and   
conservation of angular momentum.      
\citet{lamm05} assumed, based on their findings,  
that NGC\,2264 represents  
a later stage in the rotational evolution of the stars than is
shown by the ONC,  
meaning that when NGC\,2264 was at the age of the ONC, its  
period distribution  
should have been  
close to  
that observed today in the ONC. Under this   
assumption, \citet{lamm05} proposed two possible explanations  
for the   
presence of the second peak ($\sim$4-5 days) in the period distribution   
of NGC\,2264:   
\begin{enumerate}  
\item  
stars with $P$$\simgt$4-5 days  
are still locked, but at a locking period shorter than that in the ONC;  
\item  
stars with $P$$\simgt$4-5 days were locked in the   
past, with a locking period similar to the locking period in the ONC, and   
were released from their disks presumably when they were at the age of   
the ONC stars.   
\end{enumerate}  
After losing their disks (i.e., at about 1~Myr, the mean age of ONC objects),   
stars of the second group  
spun up, conserving angular momentum, which resulted   
in a shift of the period distribution of NGC\,2264 toward shorter periods than are observed in the ONC distribution.  
Although about 30\% of the stars with M$>$0.25\,M$_{\odot}$ in NGC\,2264 maintain a longer  
rotation period even as they have aged from the ONC, \citet{lamm05} considered  
hypothesis 2 as the more reliable
and that a   
locking period of about 4-5 days is not trustworthy. In addition, most of the stars with   
$P$$\simgt$4-5 days  
have an $H\alpha$ index  
that is compatible with no accretion.

Recently, \citet{gregory12} studied the magnetic field of 
T\,Tauri stars and suggested that their magnetic field topologies 
are strongly dependent on the internal stellar structure. A star can begin to 
spin up even without losing its disk. If the magnetic field dipole component 
decreases, the magnetospheric radius (where the star-disk coupling occurs) moves 
inward, to values lower than the corotation radius, and the star begins to 
spin up. According to the findings of \citet{gregory12}, this apparently occurs when
the radiative core develops. Because NGC\,2264 stars have a 
variety of masses and ages, this shift of the magnetospheric radius in this scenario would 
occur at different times and would result in different locking periods, 
whose values we would not know. 
By considering that NGC\,2264 stars are older than those of the
ONC, we expect that 
the number of stars that have already developed a radiative core
and have then spun up is greater for NGC\,2264. As a result, NGC\,2264 should 
show a period distribution shifted to shorter periods than the ONC 
distribution. 
The arguments based on magnetic field topology evolution, presented by
\citet{gregory12}, imply stellar spin-up, favoring hypothesis 2. 

The rotational properties of ONC and  NGC\,2264 stars are
discussed and analyzed in Sect. \ref{results}, based on our previous 
discussions.
The ONC was investigated by assuming that stars with periods greater than 
8\,days are locked to their disks (with $P_{\rm disk}$=8\,days) and
the other stars, with $P$$<$8\,days, evolved without a disk.
We decided to analyze the rotational properties of NGC\,2264   
under the assumptions  
\begin{itemize}  
\item{  
that stars with periods greater than 5\,days are locked  
to their disks (with $P_{\rm disk}$=5\,days) and  
stars  
with $P$$<$5\,days evolved without disk, and}   
\item{  
that NGC\,2264 stars were locked when they were   
at the age of the ONC stars, with   
$P_{\rm disk}$=8\,days, and since then they evolved, conserving angular  
momentum.}   
\end{itemize}  

\section{Results}\label{results}  
\subsection{ONC stars}\label{oncstars}  
  
The Orion Nebula cluster is the massive star formation 
region closest to us, located at 470~pc from the Sun \citep{genzel81}. 
To study the angular momentum evolution in  
pMS phase, we compared  
our  
evolutionary track sets with observational data of the ONC stars.  
The ONC data we used have been kindly provided by\ Keivan Stassun, who has widely  
worked on the rotational properties of the ONC \citep{stassun99,stassun04}. The data  
consist of  
rotation periods (\citealp{stassun99} and \citealp{herbst02}), effective temperatures and luminosities  
\citep{hillen97}, and infrared excesses, $\Delta$[I$-$K] \citep{hillen98}.

\begin{figure}[tb]  
\centering{  
\includegraphics[width=8cm]{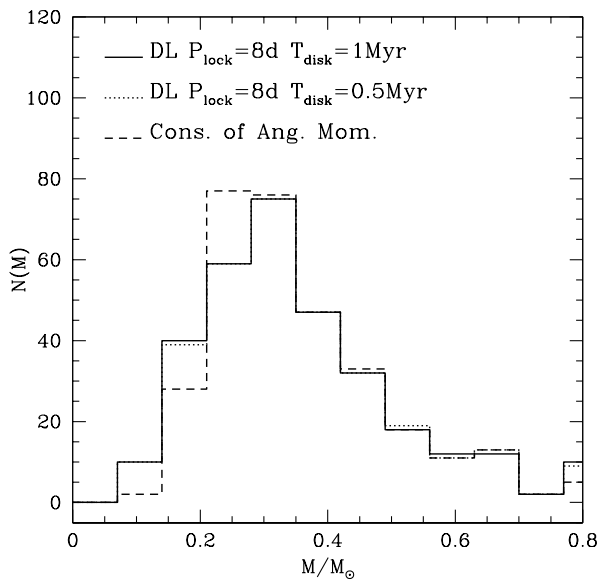}  
}  
\caption{Mass histogram of ONC stars.}  
\label{hists}  
\end{figure}

In Fig.~\ref{tracks} we show the ONC stars in the theoretical Hertzsprung-Russell
(HR) diagram
together with
our evolutionary tracks for disk-locking models  
and for models with constant angular momentum evolution.  
By using our evolutionary tracks and isochrones, we assigned a mass and an   
age to each ONC star.  
Figure~\ref{hists} shows the mass distribution of  
the observed stars, obtained by using three sets of models. Two of them  
are models with disk-locking (with P$_{\rm lock}$=8~days and   
disk lifetimes of 0.5 and 1~Myr). They produce   
very similar mass distributions, peaking at 0.3-0.35\,M$_{\odot}$.  
The third set of models, without disk-locking, produces a slightly   
different mass distribution that has a broader peak around   
0.2-0.35\,M$_{\odot}$. The resulting mass distribution of   
the bulk of the ONC population is in the range 0.2-0.4\,M$_{\odot}$,   
independently of the model used.   
  
\begin{figure}[tb]  
\centering{  
\includegraphics[width=8cm]{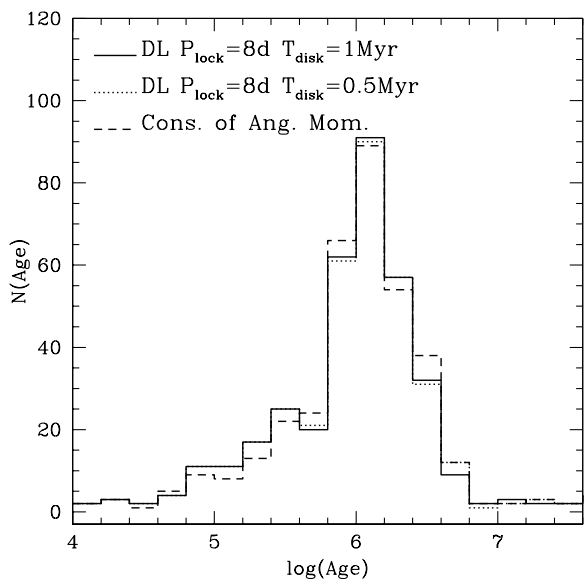}  
}  
\caption{Age histogram of ONC stars.}  
\label{hists1}  
\end{figure}  

Figure~\ref{hists1} presents the age   
distribution of our ONC star sample. The age function peaks at the same age   
($\sim$1\,Myr) for all models, and they have roughly the same age distribution.  
According to the models, the bulk of the ONC population has ages between   
0.6-2.5\ Myr, and the mean age of the stars is 1\ Myr. 

It is important to 
emphasize here that, in general, the spread in luminosities observed in the ONC and 
in other young clusters, including NGC 2264, are interpreted as a genuine age 
dispersion. However, this interpretation is questionable \citep{hartmann01}, and, 
as pointed out by \citet{jeffries11}, a coeval pMS star sample may exhibit a 
luminosity dispersion caused by observational uncertainties and physical mechanism 
that change the conventional relation between luminosity and ages. 
Consequently, Fig.~\ref{hists1} should be taken as an apparent age distribution for 
ONC stars.

\begin{figure}[t]  
\centering{  
\includegraphics[width=8.0cm]{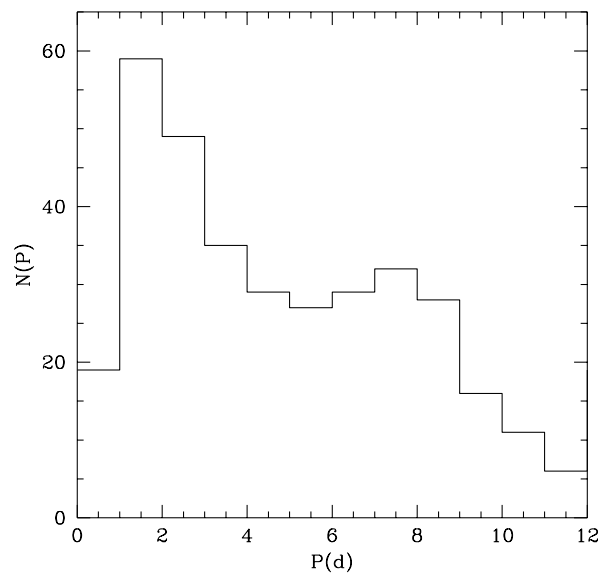}  
}  
\caption{Period histograms of all observed ONC stars \citep{landin06}.  
}  
\label{totperhistonc}  
\end{figure}  

\begin{figure}[t]  
\centering{  
\includegraphics[width=8.5cm]{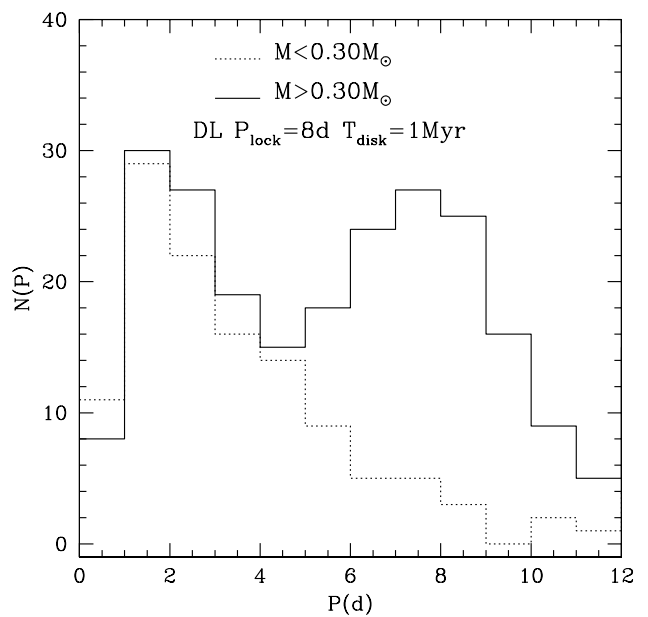}  
}  
\caption{Period histograms of the ONC stars for lower and higher mass stars.  
Dotted lines represent the objects with  
$M$$<$0.3\,M$_{\odot}$ and solid lines stand for stars with  
$M$$>$0.3\,M$_{\odot}$.}  
\label{transition}  
\end{figure}  
  
The ONC period distribution   
presents two main features: bimodality and dichotomy  
(e.g., \citealp{attridge}, \citealp{choi96}, \citealp{herbst02}, and \citealp{landin06}).  
The bimodality is  
due to its two peaks (Fig.~\ref{totperhistonc}), and  
the dichotomy, which is related to bimodality,  
is due to the dependence  
of the  
rotational properties  
on mass, that is, the different behavior of the period distribution exhibited by  
objects with masses lower and higher than a given threshold (the transition mass, $M_{\rm tr}$).  
Stars with masses higher than $M_{\rm tr}$   
present a bimodal distribution,  
while those  
with masses  
lower than $M_{\mathrm{tr}}$  
show only a tail of slow rotators.  
The value of transition mass depends on the model used to estimate the stellar  
masses. For gray models with $\alpha$=1.5 by \citet{dantona97},   
M$_{\rm tr}$=0.25\,M$_{\odot}$. For non-gray models by \citet{landin06} with   
$\alpha$=2.0, M$_{\rm tr}$$\sim$0.35\,M$_{\odot}$.  
For  
our disk-locking models ($T_{\rm disk}$=1\,Myr and $P_{\rm lock}$=8\,days),   
the transition mass that better highlights the  
dichotomy is $M_{\mathrm{tr}}$=0.3\,M$_{\odot}$. We here defined as higher mass stars
objects with M$>$0.3M$_{\odot}$ and 
as lower mass stars
objects with M$<$0.3M$_{\odot}$.   
In Fig.~\ref{transition} we show period 
histograms for ONC stars of
both
higher
and
lower mass.
It  
fully reproduces the results by \citet{herbst02}: the  
period distribution of the whole sample of ONC stars  
has two peaks,  
at periods of 2  
and 8 days, respectively.  
  
From Fig.~\ref{transition} we  
note that 67\% of the lower mass objects  
(M$<$0.3\,M$_{\odot}$)  
have $P$$<$4\,days and 14\%   
have  
$P$$>$6\,days.   
Of the higher mass stars, 38\% have periods shorter than 4 days  
and 48\% have periods greater than 6 days.   

We now establish a disk-locking criterion based on rotation 
periods. 
Stars with periods
longer
than a given threshold ($P_{\mathrm{thresh}}$=8 days for the
ONC)   
are considered to be still locked  
to a disk \citep[as suggested by][]{herbst02}.  
For stars with $P$$<$$P_{\mathrm{thresh}}$  
(which is currently considered as unlocked),  
we  
determined the epoch at which their period  
was equal to 8\,days. This would  
have been the time at which the stars would have lost their disks and, from  
then on, evolved with constant angular momentum. Following this approach, we found  
some stars that had $P$=8\,days at an age younger than $10^5$\,yr.  
Since they have lost their disk very early, these objects were considered  
to have evolved without a disk.  
A similar criterion was used by \citet{landin06}. 
\begin{figure}[t]  
\centering{  
\includegraphics[width=9.2cm]{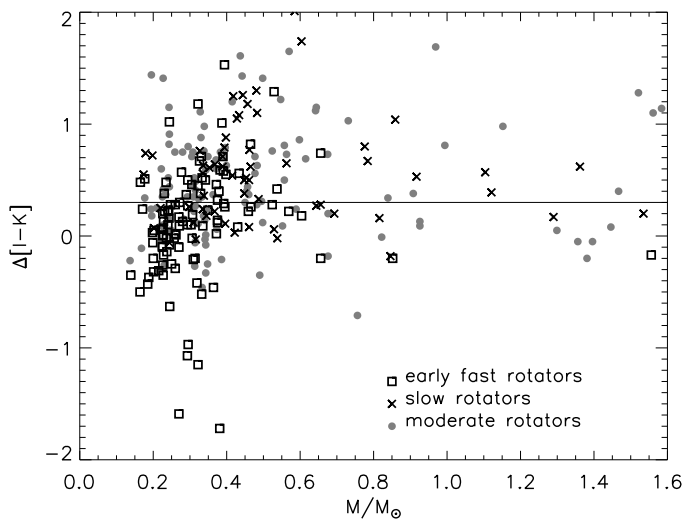}  
}  
\caption{$\Delta[I$$-$$K]$ versus stellar mass   
for early fast ($\square$ symbols),  
slow ($\times$) and moderate  
(\textcolor{gray}{$\bullet$}) rotators  
in the ONC. Figure adapted from \citet{landin06}.
}  
\label{excess}  
\end{figure}  
In this way,  
we identified three distinct populations:   

\vspace{0.2cm}
\noindent
i) early fast rotators -- stars locked only for ages $<$10$^5$\,yr; \newline   
ii) slow rotators -- stars locked to their disks ($P$$\geq$8\,d); \newline  
iii) moderate rotators -- unlocked stars ($P$$<$8\,d and ages $>$$10^5$\,yr).  
  
\vspace{0.2cm}

When we applied this criterion  
to our sample of 359 stars, 148 were classified as moderate rotators,  
76 as slow rotators, 115 as early fast rotators, and 20 stars are younger than  
10$^5$\,yr. Stars of this last group 
did not enter
our analysis.   
  
Since the near-infrared excess $\Delta[I$$-$$K]$ is an observational indicator of the presence of disks, we compared our results with the near-infrared excess measurements for 324 of our objects.
\citet{landin06} concluded that both methods agree well for slow and early fast rotators.  
IR excess provides a good diagnostic for passive disks,  
that is, for circumstellar disks not necessarily related to accreting processes.   
Still locked stars are expected to have   
$\Delta[I$$-$$K]$$>$0.3,  
while those  
evolving without a disk should have   
infrared excesses significantly
lower
than this threshold value   
\citep{herbst02}. In Fig.\,\ref{excess} we 
reproduce Fig.~8 of 
\citet{landin06} including the moderate rotators, showing   
$\Delta[I$$-$$K]$ as a function of stellar mass. 
From these, we 
found that 64\% of slow rotators have
$\Delta[I$$-$$K]$$\geq$0.3, while 70\% of early fast rotators and 42\% of moderate rotators have
$\Delta[I$$-$$K]$$<$0.3.

\begin{figure}[t]  
\centering{  
\includegraphics[width=8.0cm]{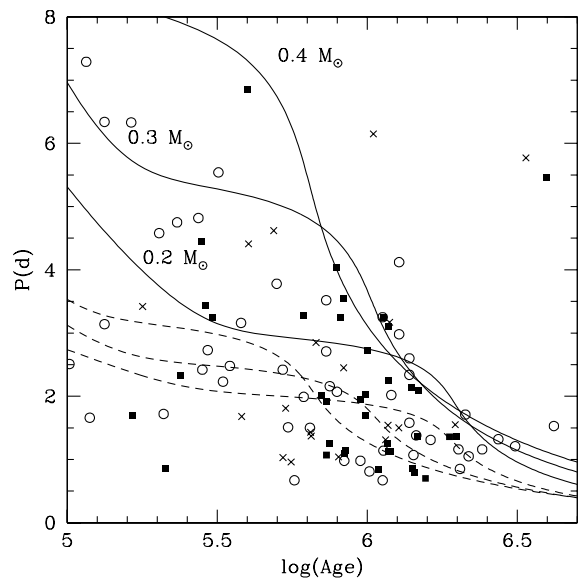}  
}  
\caption{Period  
versus age for early fast rotator  
pMS stars of the ONC.  
Their positions in the period-age plane  
are compared with models assuming   
evolution with constant angular momentum \citep{landin06}.  
Masses in the interval 0.2$<$M/M$_{\odot}$$<$0.3  
correspond to $\circ$, 0.3$<$M/M$_{\odot}$$<$0.4 to $\bull$, and  
all the remaining to $\times$.  
Solid lines represent models with $J_{\rm in}$=$J_{\rm kaw}$, while dashed lines stand for models with   
$J_{\rm in}$=$3J_{\rm kaw}$ for $M$$=$ $0.2, 0.3$ and $0.4~M_{\odot}$. Figure adapted from \citet{landin06}. 
}  
\label{evol}  
\end{figure}

According to \citet{landin06}, the evolution of the early fast rotators   
is consistent with an evolution  
conserving angular momentum from the beginning.  
Figure~\ref{evol}, taken from \citet{landin06},  
shows the ONC stars   
classified as early fast rotators in  
three mass intervals in the period-age plane, as well as
their
models  
corresponding to constant angular momentum
evolution.   
To fully bracket the observed periods, it is necessary to assume a distribution  
of initial angular momenta $J_{\mathrm{in}}$ at least in the range  
$J_{\mathrm{kaw}}$$<$$J_{\mathrm{in}}$$<$$3J_{\mathrm{kaw}}$.  
$J_{\mathrm{kaw}}$  
is described by the prescription of \citet{kawaler87} and  
in Eq.\,(\ref{kaweq}).  
We used in this work  
models with $\alpha$=2, and  
Fig.~\ref{evol} (and also  
Fig.~\ref{ngcevol}) shows that even  
models with   
$J_{\rm in}$=$3J_{\rm kaw}$ are unable to fit some  
fast rotators ($P$$\simlt$2\,days). However, models with lower convection  
efficiency ($\alpha$=1) can reproduce the period evolution of these objects along time  
\citep{landin06}.   
According to \citet{barnes01}, these fast rotators can be
fitted by solid body models with an initial rotation period of 4 days (at
birthline) and 
conserving angular momentum from the beginning.
  
The rotational evolution of slow and moderate rotators is
not consistent with  
an evolution that conserves the angular momentum since the early phases. If our criterion  
is correct, slow rotators cannot evolve conserving angular momentum because they are still   
locked to their disks. Moderate rotators experience evolution with conservation of   
angular momentum only after going through the locking phase.  
To test our disk-locking models, which are based on  
the hypothesis that a mechanism preventing the stellar spin-up exists (for a while),   
we therefore investigated first the moderate rotators in the ONC.  
According to our criterion, the moderate rotators are not 
locked to their disks anymore (i.e., they have already lost their disks). Therefore, no significant  
observational evidence of (mainly active) disk is expected.  
  
Of the 148 moderate rotators of our sample, 130 objects 
have near-infrared excess and 40 have mid-infrared measurements. Of the
130 objects of the first group, a considerable amount (58\%) have $\Delta[I$$-$$K]$$>$0.3, as shown in Fig.~\ref{excess}. 
As they are assumed to be stars that have lost
their disks, we expected a lower fraction of stars with indexes indicating
the presence of disks.  
According to our estimates of age and  
of the time at which the disks were lost,   
the majority (60\%) of the ONC stars with infrared excess   
compatible with locked stars have lost their disks relatively recently 
(less than $10^6$~yr ago).  
The reason that most moderate rotators have  
$\Delta[I$$-$$K]$$>$0.3 might be that  
they probably  
did not have time to completely  
lose the surrounding material. An alternative explanation could
be that they are still locked 
to their disks, but at a radius smaller than the corotation radius. As the
radius of the star-disk interaction moves inward when the radiative core develops 
\citep{gregory12}, stars with masses greater than the transition to complete 
convection ($\sim$ 0.3\,M$_{\odot}$) may have been classified as moderate 
rotators, but show $\Delta[I$$-$$K]$ compatible with disk presence.

\begin{figure}[t]  
\centering{  
\includegraphics[width=8.0cm]{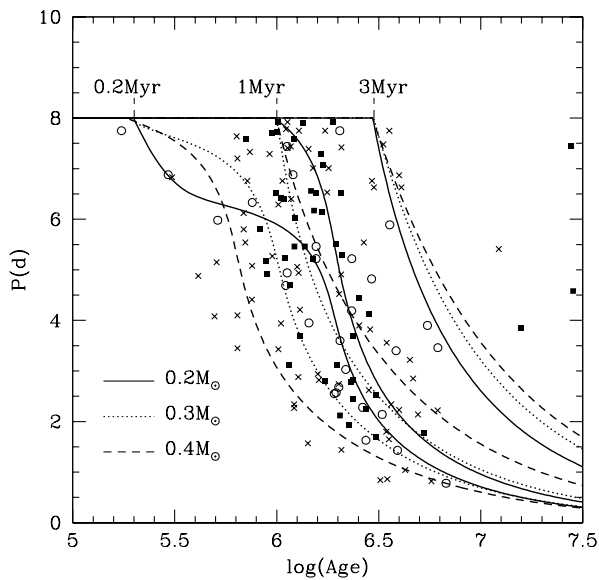}  
}  
\caption{Period  
versus age for moderate rotators  
pMS stars of the ONC.  
Their positions in the period-age plane are compared with models with    
disk-locking for three different locking times   
($T_{\rm disk}$=0.2, 1, and 3~Myr) and  
$P_{\rm disk}$=8\,days.  
The symbols are the same as in Fig.~\ref{evol},  
and the lines represent our models for 0.2 (solid), 0.3 (dotted), and  
0.4\,M$_{\odot}$ (dashed).}  
\label{evolmod}  
\end{figure}

Figure\,\ref{evolmod}  
shows the ONC moderate rotators   
in the $P_{\rm rot}$ vs.\ inferred age plane and also   
displays the results of our theoretical models, assuming that   
disk-locking is   
the mechanism preventing the stellar spinning up in the beginning of the   
pMS phase.  
According to our models, stars with 
0.2,  
0.3,  
and 0.4\,M$_{\odot}$  
represent the bulk of the ONC   
population (see Fig.~\ref{hists}), and  
our theoretical curves  
reproduce the observed \textit{\textup{loci}} of the ONC moderate
rotators.
The full distribution is well reproduced  
by using disk lifetimes between 0.2--3\,Myr, which are good estimates  
for disk survival times.  
This is supported by  
\citet{bouvier07}, \citet{barnes01}, and \citet{irwin09}.  
According to \citet{bouvier07}, 
40-60\% of the stars are still surrounded by circumstellar disks at  
2\ Myr, and this fraction decreases to  
10-25\% at 10\ Myr.   
By using disk-locking models  
for solar-type stars (0.6-1.2\,M$_{\odot}$), \citet{barnes01} and 
\citet{irwin09} reproduced the rotation period of stars that
we classified as 
moderate rotators by using disk lifetimes similar to ours. The
differentially rotating models of \citet{barnes01} required locking times of 
0.3-1~Myr and initial rotation periods at birth line in the range of 4-16~days.
The solid-body models of \citet{irwin09} required locking times of 2-5~Myr and
initial rotation periods of around 8-16~days. For the low-mass domain 
(0.1-0.4\,M$_{\odot}$), solid-body models of \citet{irwin09} reproduced the ONC data
with disk lifetimes of 4-8~Myr and initial rotation periods of
around 8-23~days. These long initial periods (or P$_{\rm lock}$) are necessary to account for 
the slowest rotators, in this work called slow rotators. 
  
\subsection{NGC\,2264 stars - hypothesis 1}\label{ngc2264}  
NGC\,2264  
is a fairly populous nearby cluster,  
located at $\sim$760pc from the Sun \citep{sung97}.  
Our sample  
consists of 405 objects classified as periodic  
variables by \citet{lamm04},  
most of which  
probably are in the  
pMS   
phase. Most of the observational data of NGC\,2264 stars used in this work (rotation periods, colors, 
and $H_{\alpha}$ based accretion disk diagnostic) come from \citet{lamm05}.  
We used both the ground-based periods determined by \citet{lamm05} 
and the space-based CoRoT periods obtained by \citet{affer13}, despite the fact that the number of stars whose
periods may have been erroneously determined by the ground-based method represents only
3\% of our sample.
Since we  
lack information about  
their $T_\mathrm{eff}$ and  
luminosities  
(except for a small sample of 109 objects by \citealt{flaccomio06} and 
256 objects by \citealt{venuti14}),  
we cannot show 
them all in the classical HR   
diagram. Instead, we have  
magnitudes and colors in the Johnson V and Cousins  
R, I systems. To compare our theoretical evolutionary tracks with   
observational data, we used color-temperature relations and bolometric corrections  
by \citet{vandenberg03}. As individual reddening and extinction are not available  
for a significant number of NGC\,2264 stars, we followed \citet{lamm05} and   
calculated reddened isochrones and evolutionary tracks by using average values   
of reddening, $E(R_c$$-$$I_c)$=0.10$\pm$0.02, and extinction, $A_{I_c}$=0.25,   
toward NGC\,2264. These values were determined by \citet{rebull02} adopting  
$R$=$E(B$$-$$V)/A_v$=3.1.   
  
\begin{figure}[t]  
\centering{  
\includegraphics[width=8cm]{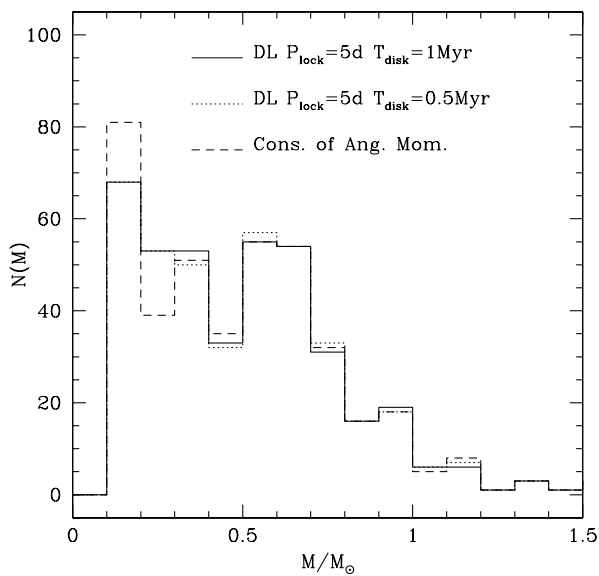}  
}  
\caption{Mass histogram of NGC\,2264 stars.  
}  
\label{ngchistm}  
\end{figure}

We estimated  
mass and  
age to each NGC\,2264 star by using our tracks and  
isochrones.   
Figure\,\ref{ngctracks} shows our models (with conservation of  
angular momentum and disk-locking for P$_{\rm lock}$=5~days and T$_{\rm disk}$=1~Myr)  
and NGC\,2264 data in the  
color-magnitude diagram $I_c$ vs.\ $R_c$$-$$I_c$.  
Figure\,\ref{ngchistm}  
shows the mass distribution of the   
observed stars obtained with models with conservation of angular momentum and   
disk-locking for P$_{\rm lock}$=5~days (T$_{\rm disk}$=1 and 0.5~Myr). 
For masses higher than $0.4\,M_{\odot}$ the three 
distributions are quite similar, showing only small   
differences. However, for masses lower than $0.4\,M_{\odot}$ they exhibit
appreciable differences that are more pronounced for models with 
conservation of angular momentum.
The mass distributions obtained with our models   
peak in the 0.1-0.2\,M$_{\odot}$ range. 
A secondary peak is seen in the   
0.5-0.6\,M$_{\odot}$ range, while the  
resulting mass distribution of the bulk of  
NGC\,2264 population is in  
the interval of 0.1-0.6\,M$_{\odot}$. 
  
\begin{figure}  
\centering{  
\includegraphics[width=8cm]{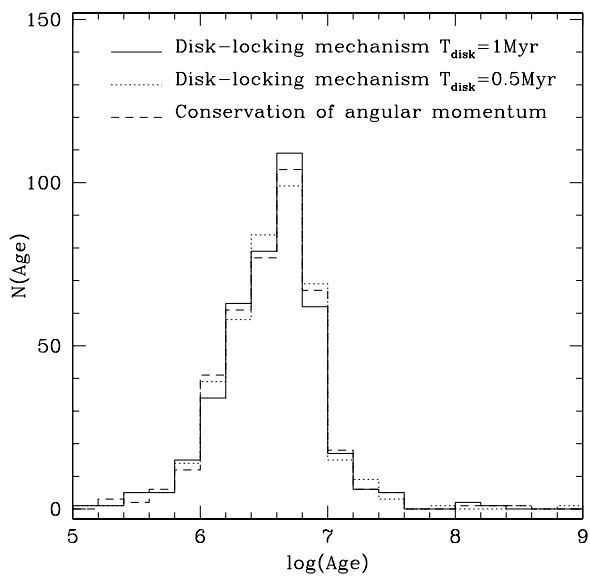}  
}  
\caption{Age histogram of NGC\,2264 stars.  
}  
\label{ngchistage}  
\end{figure}

Figure \ref{ngchistage}  
displays the apparent age distribution  
obtained with our models for this sample of   
NGC\,2264 stars, showing that  
the ages predicted by the  
three models are roughly similar, with only minor differences.  
The age function peaks in the range of 3.9-6.3\,Myr  
and the bulk of the 
population  
lies in the interval of 1-10\,Myr.   
The mean age of NGC\,2264 obtained with our three models is 3.5\,Myr.  
  
\begin{figure*}[t]  
\centering{  
\includegraphics[width=8.3cm]{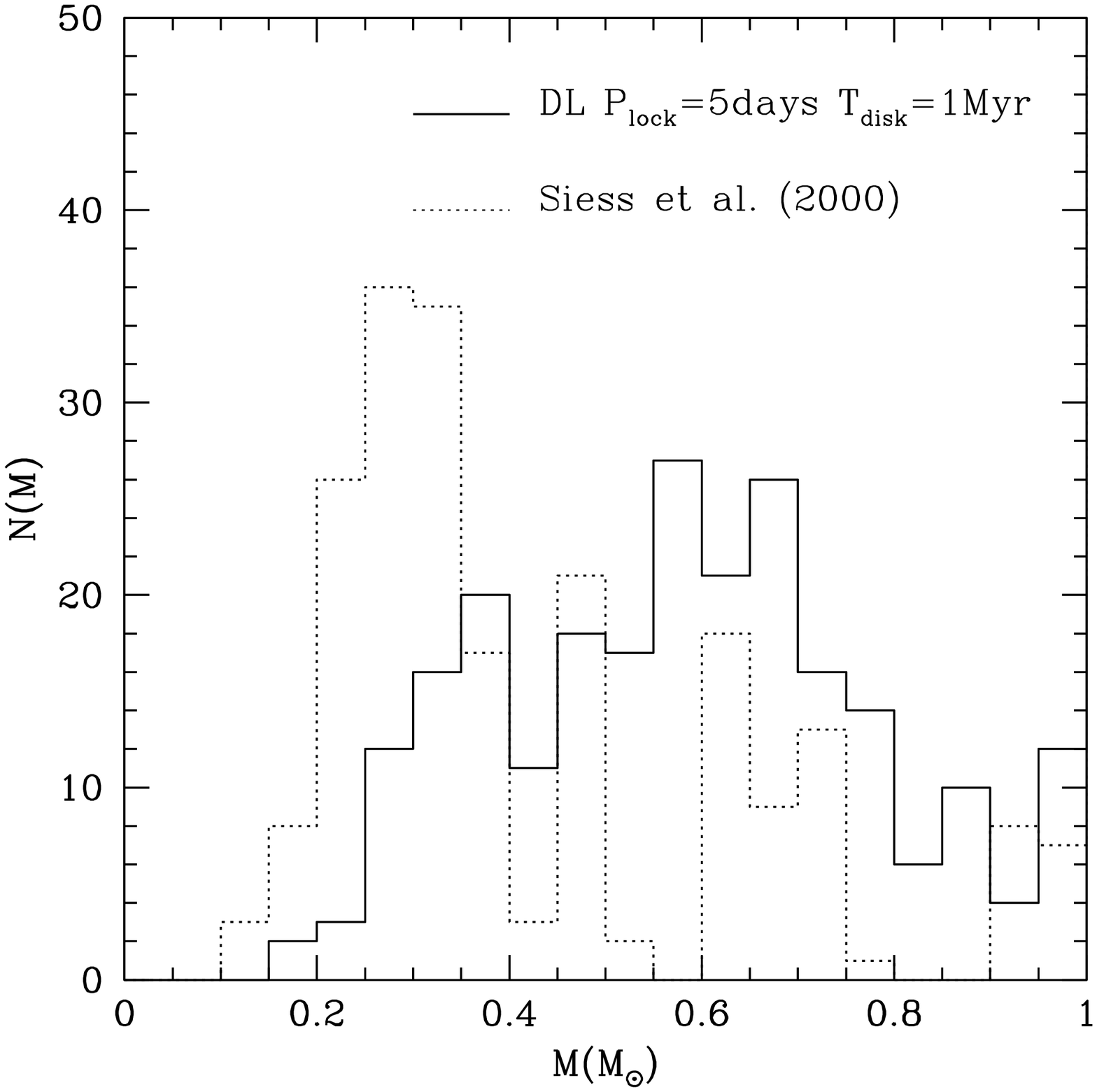}  
\includegraphics[width=8.3cm]{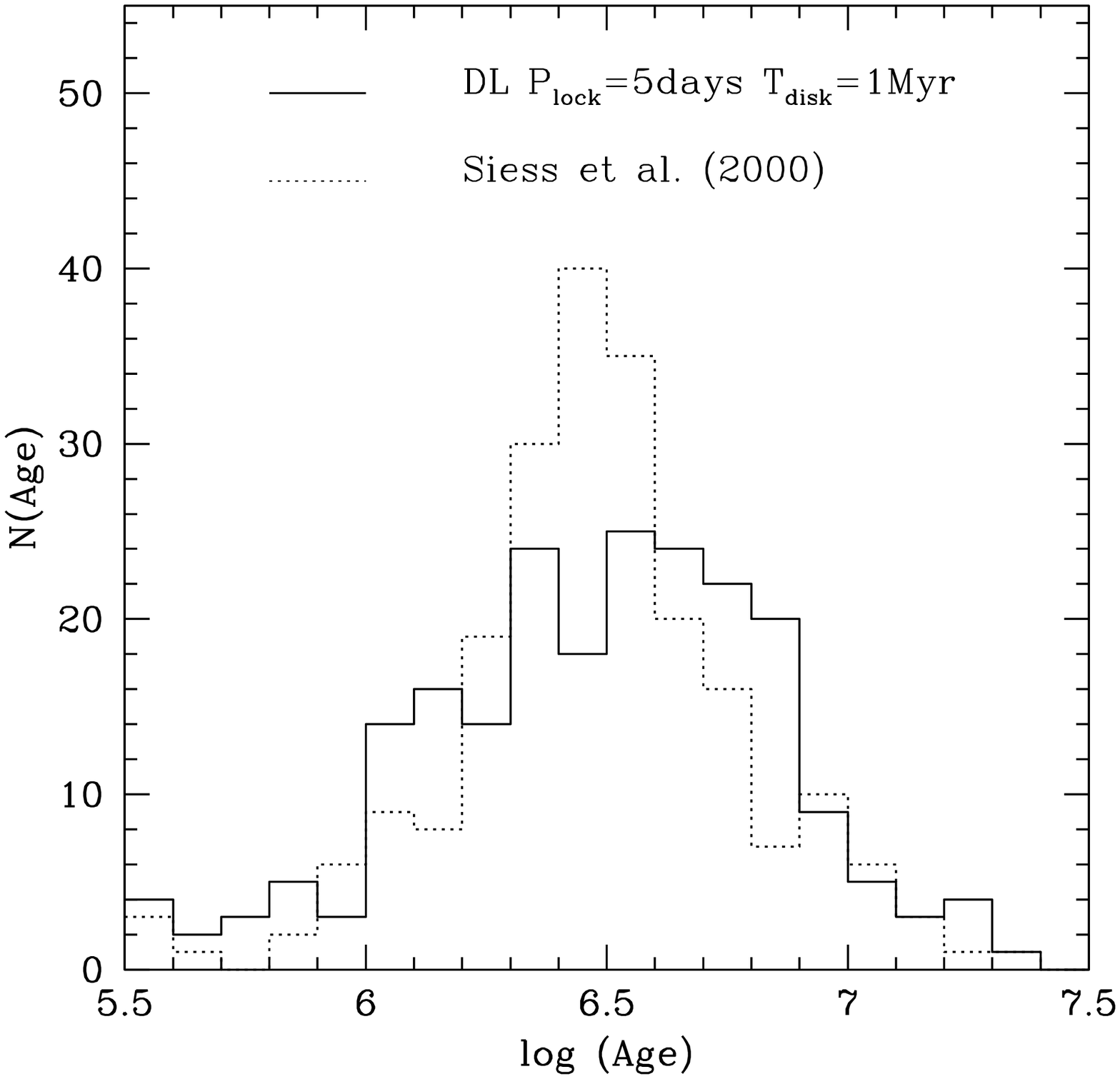}  
}  
\caption{Mass (left) and age (right)  distributions for NGC\,2264 stars in common in our sample and that of \citet{venuti14}. Solid curves 
refer to masses and ages obtained with our models, dotted curves refer 
to masses and ages obtained with the models of \citet{siess00}.  
}  
\label{compmassageven}  
\end{figure*}

\citet{venuti14} also determined stellar masses for 256 and ages for 218 
NGC\,2264 objects. 
The mean difference found for masses between their values and ours is 0.2~M$_{\odot}$, with 59\% of the 
objects having differences in mass determinations smaller than 0.2~M$_{\odot}$. 
The ages of these stars, estimated by both works, are between 1 and 10~Myr,
but the age distribution found by \citet{venuti14} has a narrow peak around 
3.2~Myr, while our age determinations are mainly distributed in the range of 2.5 
and 8~Myr (see right panel of Fig.~\ref{compmassageven}). One source of
the differences in mass and age estimates is the different 
evolutionary models used in each work. 
\citet{venuti14} used models of \citet{siess00}, while we used the non-gray models
described in Sect.~\ref{models}, which are more suitable for p-MS than the gray
atmospheric boundary conditions used by \citet{siess00}. 
The use of non-gray models mainly affects mass determinations of low-mass stars. 
The left panel of Fig.~\ref{compmassageven} shows that the mass distribution of 
NGC 2264 stars obtained with our non-gray models (solid curve) is shifted 
to higher masses in comparison with the mass distribution determined by 
\citet[dotted curve]{venuti14}. Non-gray models yield evolutionary tracks cooler 
than gray models and, as a result, mass estimates obtained from the latter are 
systematically (and erroneously) lower than those obtained from the former. 

\begin{figure}  
\centering{  
\includegraphics[width=8cm]{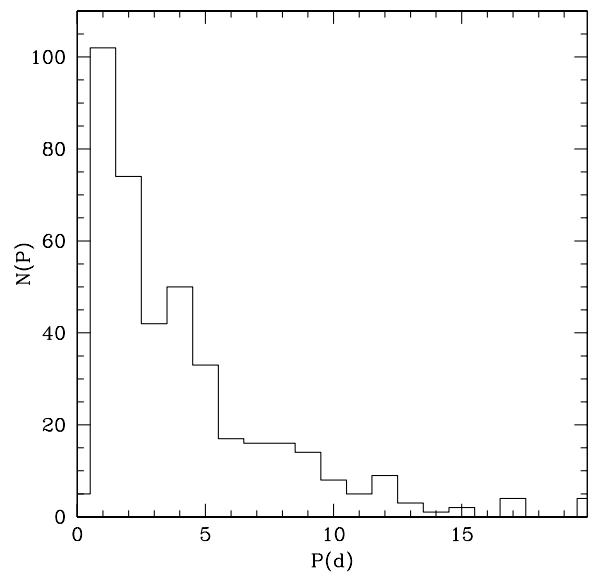}  
}  
\caption{Total period distribution for NGC\,2264 stars.  
}  
\label{distpertot}  
\end{figure}

\begin{figure*}[t]  
\centering{  
\includegraphics[width=8.3cm]{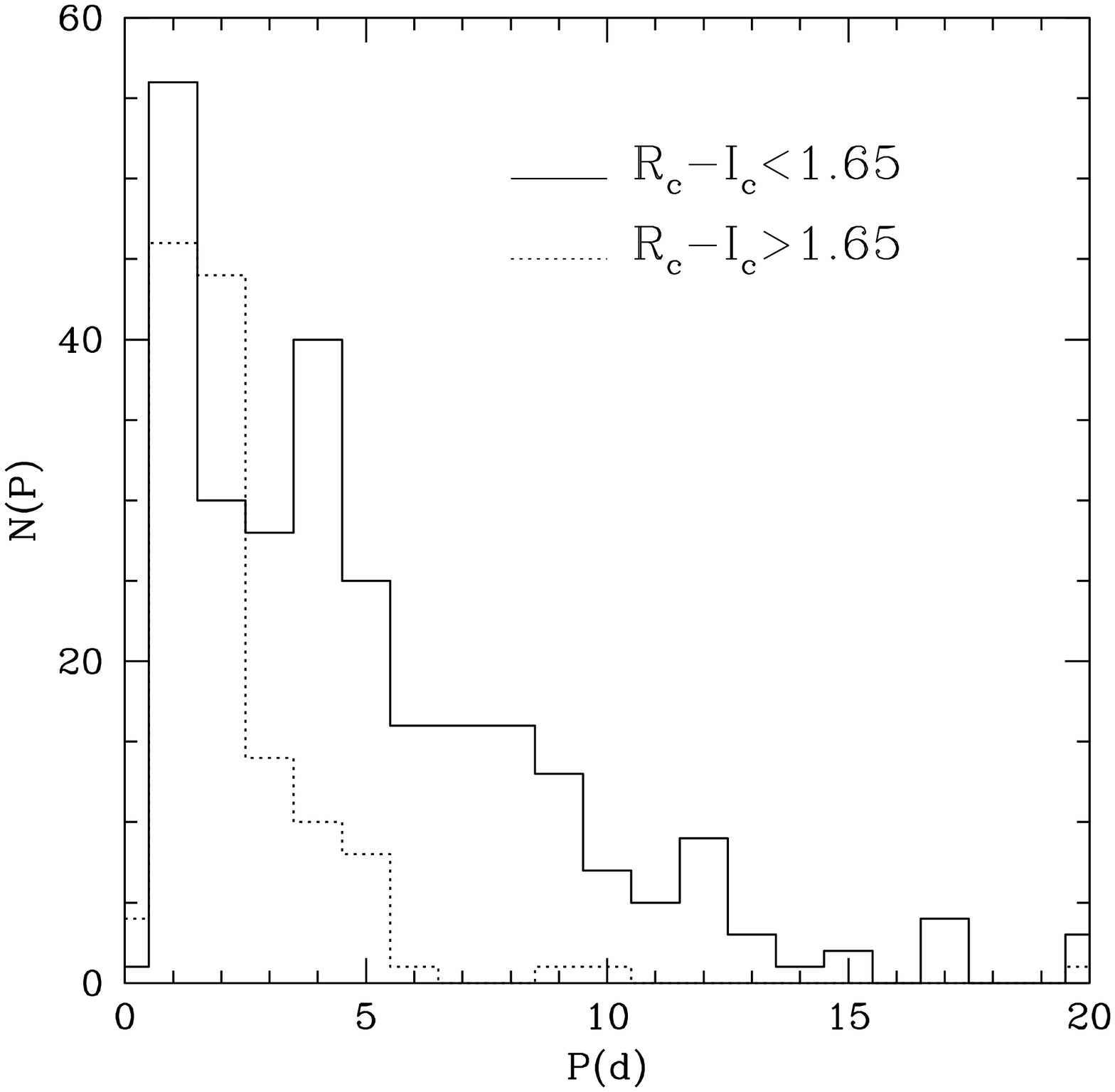}  
\includegraphics[width=8.3cm]{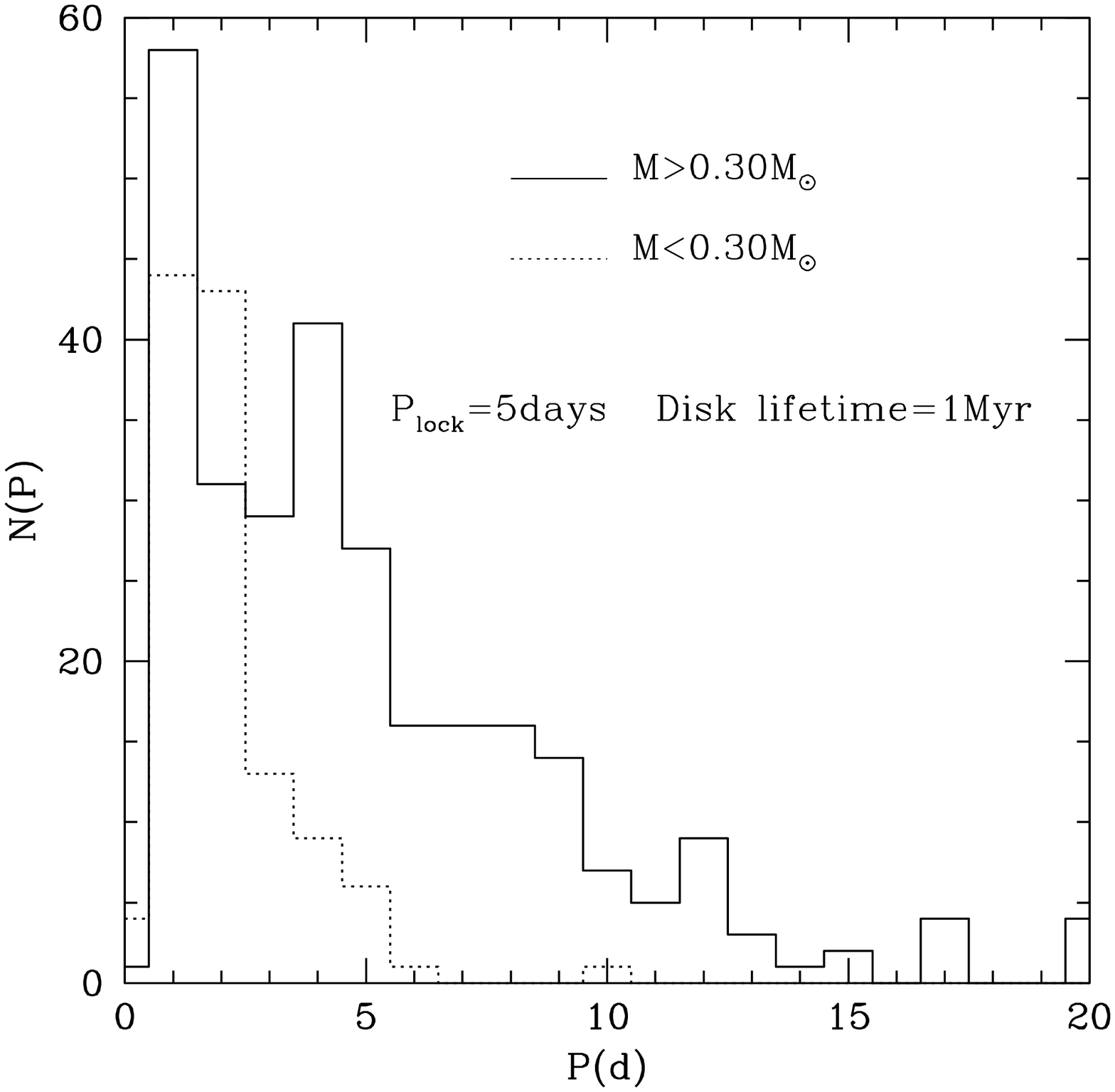}  
}  
\caption{Period histograms for our sample of pMS stars   
of NGC\,2264. The left panel shows the period distribution for stars with   
$R_c$$-$$I_c$$<$1.65 mag  
(solid line) and  
with $R_c$$-$$I_c$$>$1.65 mag (dotted line). This corresponds   
approximately to a division into stars with $M$$>$0.3\,M$_{\odot}$  
(solid line) and   
$M$$<$0.3\,M$_{\odot}$  
(dotted line),  
in the right panel.  
}  
\label{ngcdicho}  
\end{figure*}

As  
for the ONC, the rotation period distribution of NGC\,2264 also  
exhibits features such as  
bimodality and dichotomy (\citealp{lamm05} and \citealp{cieza07}), as shown in  
the  
period  
histograms  
for  
NGC\,2264 stars,  
Figs.~\ref{distpertot} and \ref{ngcdicho}. The total period distribution 
(Fig.~\ref{distpertot}) is  
bimodal, with a primary peak  
at $\sim$1\,day and a secondary one  
at $\sim$4-5\,days.  
The distribution is clearly shifted toward  
lower periods than in  
the ONC one   
(see Fig.~\ref{totperhistonc}).  
  
As  
shown in   
Fig.\,\ref{ngcdicho}, NGC\,2264 stars can be divided into two distinct populations:  
(i) a group  
with a unimodal period distribution, with a peak  
$\sim$1-2 days and a tail of slow rotators, and  
(ii) another group  
with a   
bimodal  
distribution, with  
peaks at  
$\sim$1\,d and  
$\sim$4-5\,days. In the left panel of Fig.~\ref{ngcdicho}, these groups are divided into stars with   
$R_c$$-$$I_c$$>$1.65 mag\footnote{We chose a transition  
color index of 1.65 mag, which most clearly shows the phenomenon of   
dichotomy. On the other hand, \citet{lamm05} used a value of 1.3 mag, which  
corresponds to the same transition mass adopted for the ONC by using gray models,   
i.e., 0.25~M$_{\odot}$. As can be seen in their Fig. 3a, the peak of the lower mass stars is more than twice as large as the
peak of higher mass stars in the period  
distributions.}  
and  
with  
$R_c$$-$$I_c$$<$1.65 mag.  
This transition color index corresponds to  
0.3\,M$_{\odot}$, according to our disk-locking models  
($T_\mathrm{disk}$=1\,Myr and   
$P_{\rm lock}$=5\,d). We chose 
this value for the transition mass  
to better highlight the dichotomy.   
However, a dichotomy is observed for a wide range of   
transition masses (0.2$\leq$M$_{\rm tr}$/M$_{\odot}$$\leq$0.6, for NGC\,2264), with different period   
distributions for each transition mass. For M$_{\rm tr}$$\sim$0.35\,M$_{\odot}$,  
the peak of low-mass stars and the primary peak of high-mass stars have nearly 
the same height.  
The right panel of   
Fig.~\ref{ngcdicho} shows  
NGC\,2264 objects divided into  
groups with masses greater  
and  
lower than 0.3\,M$_{\odot}$.  
This value of the transition mass is the same  
for models  
with either  
conservation of angular momentum  
or  
disk-locking, although the distributions are slightly different. However,   
$M_{\rm tr}$ is  
sensitive to other parameters such as atmospheric   
boundary conditions and convection treatment \citep{landin06}.   
Figure\,\ref{ngcdicho}  
gives a hint on how rotational properties of  
NGC\,2264 behave as a function of mass.  
Of our sample of 405 stars, we estimated that 280 stars  
have M$>$0.3~M$_{\odot}$ and 125 have M$<$0.3~M$_{\odot}$. Of the objects with  
M$<$0.3~M$_{\odot}$ 72.8\% have P$<$2.5~days and 16.8\% have P$>$3.5~days.   
Of the objects with M$>$0.3~M$_{\odot}$ 32.1\% have P$<$2.5~days and   
57.9\% have P$>$3.5~days.  
  
We used the same  
method  
as described in Sect.~\ref{oncstars}   
to establish a disk-locking criterion based on the  
rotation period.  
For NGC\,2264 we  
use in this section another value for  
the threshold period because the secondary peak in the period distribution  
for higher mass stars occurs around 4-5\,days  
(instead of 8\,days, as in the ONC).  
In the analysis of the rotational history of  
NGC\,2264,  
we  
use  
$P_{\rm thresh}$=5\ days   
and we have, again, three distinct groups:  

\vspace{0.20cm}

\noindent
i) early fast rotators -- stars locked only for ages $<$10$^5$\,yr; \newline   
ii) slow rotators -- stars probably disk locked (P$\geq$5\ d); \newline
iii) moderate rotators -- unlocked stars ($P$$<$5\,d and ages $>$$10^5$\,yr).

\vspace{0.20cm}

\noindent
The sample of \citet{lamm05} with its 405 stars has 240
that were found to be moderate rotators,
117 are slow
and 48 early fast rotators. 

We tested this disk-locking criterion against observational indicators of disk presence 
that are related, mainly, with $H\alpha$ emission (H$\alpha$-index, 
$\Delta[R_c$$-$$H\alpha]$) by 
\citet{lamm03} and \citet{lamm04}. These authors used nearly 
simultaneous observations in $V$, $R_C$, $I_C,$ and $H_{\alpha}$, obtained 
between December 2000 and March 2001, to search for rotation 
periods and to
determine H$\alpha$-indexes. As chromospheric activity can also
produce considerable $H\alpha$ emission, its presence is an ambiguous
sign of accretion disk, in particular close to an $H\alpha$ emission equivalent width
of $W_{\lambda}(H\alpha)$=10$\AA$ \citep{lamm03}. As complementary data,
we therefore used {\it Spitzer} IRAC data kindly provided by Paula Teixeira 
\citep{teixeira12}, who classified the circumstellar disks of
NGC\,2264 stars based on their spectral energy distribution slope between 
3.6$~\mu m$ and 8$~\mu m$, defined as $\alpha_{IRAC}$. These
IRAC data were obtained in two epochs, March and October 2004, and can help us 
to distinguish the stars with a disk from diskless objects. 
In addition, we investigated 
$W_{\lambda}(H\alpha)$, $U$$-$$V$ and near-infrared excesses ($I$$-$$K$
and $H$$-$$K$) by 
\citet{rebull02}, near-infrared excess and $W_{\lambda}(H\alpha)$ by 
\citet{dahm05}, mid-infrared excess by \citet{cieza07}, and {\it u}-band
exccess by \citet{venuti14}. For disk 
identification purposes, we used the following values to indicate 
the presence of a disk: 
$W_{\lambda}(H\alpha)$$\ge$10$\AA$, $\Delta[U$$-$$V]$$\le$$-$0.5,
$\Delta[I$$-$$K]$$\ge$0.3, $\Delta[H$$-$$K]$$\ge$0.15, 
$[3.8$$-$$8.0]$$\ge$0.7, and $\alpha_{IRAC}$$>$$-$2.56.

As in \citet{lamm04,lamm05}, we also dealt with the $H\alpha$ index, 
$\Delta[R_c$$-$$H\alpha]$, 
which is a measure of the $H\alpha$ emission. The $H\alpha$ index is 
defined by the following equation: 
\begin{equation}
\Delta[R_c-H\alpha]=(R_c-H\alpha)_{\rm star}-(R_c-H\alpha)_{\rm locus},
\label{haindex}
\end{equation}
\noindent where $(R_c$$-$$H\alpha)_{\rm star}$ is the instrumental color of the 
star and $(R_c$$-$$H\alpha)_{\rm locus}$ is the median $(R_c$$-$$H\alpha)$ color
as a function of $R_c$$-$$I_c$ for the locus of pMS/MS stars. We
obtained $(R_c$$-$$H\alpha)_{\rm locus}$ for our sample by using the
equation \citep{lamm04}
\begin{equation}
(R_c\! -\!H\alpha)_{\rm locus}\!=\!-0.06(R_c\!-\!I_c)^2\!+\!0.38(R_c\!-\!I_c)\!-\!3.50,
\label{rhaloc}
\end{equation}
\noindent which should be valid for an MS star. 

\begin{figure}[h]  
\centering{  
\includegraphics[width=9.10cm]{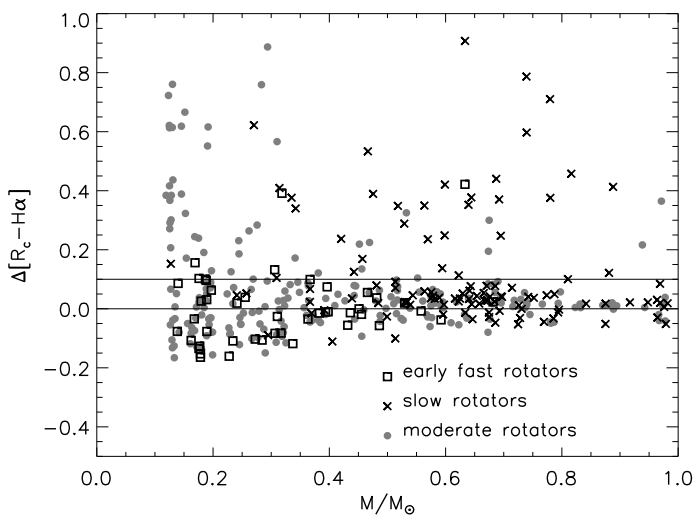}  
}  
\caption{$\Delta[R_c-H\alpha]$ versus stellar mass for NGC\,2264 stars  
classified as early fast rotators  
($\square$), slow rotators   
($\times$),  
and moderate rotators  
(\textcolor{gray}{$\bullet$}).  
}  
\label{drmha}  
\end{figure}

To relate this $H\alpha$ index to some physical property,  
\citet{lamm03} and \citet{lamm04} used stars with a known $H\alpha$ equivalent width,   
$W_{\lambda}(H\alpha)$,  
to compare these two quantities. They found   
that 83\% of the stars with $\Delta[R_c$$-$$H\alpha]$$\geq$0.1 mag have   
$W_{\lambda}(H\alpha)$$\simgt$10\,{\AA} and are likely \textsc{ctts}.  
On the other hand, about 85\% of the stars with $\Delta[R_c$$-$$H\alpha]$$<$0.1   
mag have $W_{\lambda}(H\alpha)$$<$10\,{\AA} and are most likely \textsc{wtts}.  
In Fig.~\ref{drmha} we show $\Delta[R_c$$-$$H\alpha]$ versus stellar mass for   
NGC\,2264  
objects,  
indicating those  
classified as  
early fast,  
slow,  
and moderate   
rotators. Slow rotators are quite concentrated in
the more massive region of the plot, early fast rotators in the less
massive region. The sample of moderate rotators spreads throughout the entire mass interval. Figure~\ref{drmha} shows only objects with
$M$$\le$$1.0M_{\odot}$. However, Fig.~\ref{ngchistm} shows that only a few 
objects of our sample stay outside this range of mass.

Following \citet{lamm03},
we call stars with $\Delta[R_c$$-$$H\alpha]$$-$$\delta[R_c$$-$$H\alpha]$$\geq$0.1 \textsc{ctts} and stars with 
$\Delta[R_c$$-$$H\alpha]$$+$$\delta[R_c$$-$$H\alpha]$$\leq$0.0 \textsc{wtts}. The quantity $\delta[R_c$$-$$H\alpha]$ is the photometric
error of $R_c$$-$$H\alpha$ color. Stars that are neither \textsc{ctts}
nor \textsc{wtts} according to these criteria are called intermediate
cases. They have a small (positive) $H\alpha$-index and could be 
\textsc{ctts}. 
  
\citet{lamm04}  
noted that their periodic variables (which form our sample of stars in   
NGC2264) are biased toward the   
\textsc{wtts},  
objects with $\Delta[R_c$$-$$H\alpha]$ $<$0.1,  
mentioning that  
this is  
a limitation of the photometric method used to study  
the rotational periods among T\,Tauri stars.
In our sample of 405 stars, 81 have 
$\Delta[R_c$$-$$H\alpha]$$-$$\delta[R_c$$-$$H\alpha]$
$\geq$0.1 (\textsc{ctts}), 
114 stars have
$\Delta[R_c$$-$$H\alpha]$$+$$\delta[R_c$$-$$H\alpha]$
$\leq$0.0 
(\textsc{wtts}), and 210 were classified as ``intermediate cases''. 
As expected, according to the classification of \citet{lamm03},
for \textsc{ctts} and 
\textsc{wtts} $\text{the }H\alpha$-index values agree well with the
$W_{\lambda}(H\alpha)$ measurements by \citet{rebull02} and \citet{dahm05}. Of the 81
stars classified as \textsc{ctts,} 42 have a $W_{\lambda}(H\alpha)$ measure in
at least one of these two works, and 38 have $W_{\lambda}(H\alpha)$$\ge$10$\AA$. Of the 114
stars classified as \textsc{wtts,} 42 have at least one 
$W_{\lambda}(H\alpha)$ 
measurement, and 39 of them have $W_{\lambda}(H\alpha)$$<$10$\AA$. 
Of the 210 stars classified as intermediate cases, 98 have at least 
one $W_{\lambda}(H\alpha)$ measurement, and 18 have 
$W_{\lambda}(H\alpha)$$\ge$10$\AA$ and could be \textsc{ctts}.

Unfortunately, the only observational disk indicator for which 
we have measurements for all stars is $\Delta[R_c$$-$$H\alpha],$ and this indicator is inconclusive 
for more than half of our sample stars, classifying 210 objects as intermediate 
cases. To distinguish these stars as objects with and without
disks, we also used the $\alpha_{IRAC}$ index reported by \citet{teixeira12}. Of the 405 
stars of our sample, 170 have $\alpha_{IRAC}$ 
measurements, 75 have $\alpha_{IRAC}$$>$$-2.56$ 
(indicating that they have dust in the inner disk) and 95 have 
$\alpha_{IRAC}$$\leq$$-2.56$ (indicating that they have naked photospheres).
These IRAC data agree very well with the $H\alpha$ index.
Of the stars with $\alpha_{IRAC}$ measurements, 35 were classified
as \textsc{ctts} (34 have $\alpha_{IRAC}$$>$$-2.56$ and only 1 
has $\alpha_{IRAC}$$\leq$$-2.56$), 40 were classified as \textsc{wtts}
(32 have $\alpha_{IRAC}$$\leq$$-2.56$ and 8 have $\alpha_{IRAC}$$>$$-2.56$), and 95 were classified as intermediate cases (62 have 
$\alpha_{IRAC}$$\leq$$-2.56$ and 33 have $\alpha_{IRAC}$$>$$-2.56$).
By combining both indexes, $H\alpha$ and $\alpha_{IRAC}$, we identified 114 
objects with observational evidence of disk presence (\textsc{ctts} and
intermediate cases with $\alpha_{IRAC}$$>$$-2.56$), 176 without such 
indications (\textsc{wtts} and intermediate cases with 
$\alpha_{IRAC}$$\le$$-2.56$), and 115 stars for which we do not have enough 
information to conclude whether they have a disk (intermediate cases without
$\alpha_{IRAC}$ measurements). 
By comparing our disk-locking criterion and these observational
disk diagnostics, we found that only 40\% of objects with observational 
indications of disk presence are slow rotators. For those without 
observational indications of a disk,
the two methods agree well because 75.5\% of them are early fast or moderate rotators.
Of the 115 objects labeled intermediate cases that do not have 
$\alpha_{IRAC}$ measurements, 66\% are moderate rotators, 10.5\% are early 
fast rotators, and 23.5\% are slow rotators.

We can also compare the $H\alpha$ index with other disk indicators,
such as $U$$-$$V$, $I$$-$$K$, $H$$-$$K$ excesses \citep{rebull02} and the [3.6-8.0] 
excess \citep{dahm05}.
In the \textsc{ctts} group, 43 have measurements of 
another disk indicator, 
which for 23 stars (53.5\%) indicate the presence of disks. 
In the \textsc{wtts} group, 52 stars have at least one measurement of another 
disk indicator, which for 40 stars (76.9\%) suggests that they are diskless stars.
In the group of 210 stars classified as intermediate cases, 128 have
at least another measure of a disk indicator, which indicates
for 60 the presence of a disk and that they might be \textsc{ctts}. 
This analysis reveals that we
do not have enough UV and infrared data for our star sample to 
compare them with the $H\alpha$ index. For the stars for which we have another
disk indicator measurement, the comparison between them and $H\alpha$ index 
do not agree well for stars classified as \textsc{ctts,} but they 
agree for the \textsc{wtts} stars.

%**********************TABLE****************************
\begin{table}[htb]
\footnotesize  
\caption{Comparison between our disk-locking criterion and observational data.} 
\label{tabdlockcrit}
\begin{tabular}{|l|l|l|}
\hline
P-based criterion & $\displaystyle H_{\alpha}$ emission & $\displaystyle \alpha_{IRAC}$ \\ \hline
\multicolumn{1}{ |c| }{\multirow{9}{*}{117 slow rotators} } & \multirow{2}{*}{30 \textsc{ctts}} & $\displaystyle 17\ \alpha_{IRAC}>-2.56$ \\
& & $\displaystyle 1\ \alpha_{IRAC}\leq-2.5$ \\
& & 12 no measurements \\ 
\cline{2-3}
& \multirow{2}{*}{18 \textsc{wtts}} & $\displaystyle 5\ \alpha_{IRAC}>-2.56$ \\
& & $\displaystyle 7\ \alpha_{IRAC}\leq-2.5$ \\
& & 6 no measurements \\ 
\cline{2-3}
& \multirow{2}{*}{69 Interm. cases} & $\displaystyle 17\ \alpha_{IRAC}>-2.56$ \\
& & $\displaystyle 25\ \alpha_{IRAC}\leq-2.5$ \\
& & 27 no measurements \\ \hline
\multicolumn{1}{ |c| }{\multirow{9}{*}{48 early fast rotators} } & \multirow{2}{*}{6 \textsc{ctts}} & $\displaystyle 2\ \alpha_{IRAC}>-2.56$ \\
&  & $\displaystyle 0\ \alpha_{IRAC}\leq-2.5$ \\ 
& & 4 no measurements \\
\cline{2-3}
& \multirow{2}{*}{24 \textsc{wtts}} & $\displaystyle 1\ \alpha_{IRAC}>-2.56$ \\
& & $\displaystyle 8\ \alpha_{IRAC}\leq-2.5$ \\
& & 15 no measurements \\ 
\cline{2-3}
& \multirow{2}{*}{18 Interm. cases} & $\displaystyle 1\ \alpha_{IRAC}>-2.56$ \\
& & $\displaystyle 5\ \alpha_{IRAC}\leq-2.5$ \\ 
& & 12 no measurements \\ \hline
\multicolumn{1}{ |c| }{\multirow{9}{*}{240 moderate rotators} } & \multirow{2}{*}{45 \textsc{ctts}} & $\displaystyle 15\ \alpha_{IRAC}>-2.56$ \\
& & $\displaystyle 0\ \alpha_{IRAC}\leq-2.5$ \\
& & 30 no measurements \\ 
\cline{2-3}
& \multirow{2}{*}{72 \textsc{wtts}} & $\displaystyle 2\ \alpha_{IRAC}>-2.56$ \\
& & $\displaystyle 17\ \alpha_{IRAC}\leq-2.5$ \\
& & 53 no measurements \\ 
\cline{2-3}
& \multirow{2}{*}{123 Interm. cases} & $\displaystyle 15\ \alpha_{IRAC}>-2.56$ \\
& & $\displaystyle 32\ \alpha_{IRAC}\leq-2.5$ \\ 
& & 76 no measurements \\ \hline
\end{tabular}
\end{table}
%**********************TABLE****************************

We now compare the {\it u}-band survey by \citet{venuti14} with
$H\alpha$-indexes and $\alpha_{IRAC}$ measurements and with our disk-locking
criterion.
Of the 405 stars of our sample, 256 objects have {\it u}-band 
photometric measurements, 59 objects
have {\it u}-band excess (compatible with ongoing accretion), and 197 stars
have no {\it u}-band excess. Of the 59 stars
with {\it u}-band excess, 31 are classified as \textsc{ctts}, 6 as 
\textsc{wtts,} and 22 as intermediate case stars. Of the 197 objects 
without {\it u}-band excess, 16 are classified as \textsc{ctts}, 62 as 
\textsc{wtts,} and 119 as intermediate case objects. 
Although {\it u}-band photometry data have been obtained on December 2010,
some years after $\Delta[R_c-H_{\alpha}]$ data, they agree reasonably well 
as accretion diagnostics. In addition, all 170 objects of our sample that have
$\alpha_{IRAC}$ measurements are present in the {\it u}-band survey, and the two 
diagnostics agree \citep{venuti14}. In this subsample, 89 stars have 
no {\it u}-band excess and $\alpha_{IRAC}$$\leq$$-$2.56 (no dust in the inner disk),
40 objects do not show either {\it u}-band excess or an IRAC inner disk, 
and only 6 objects show {\it u}-band excess and do not have dusty inner disk. 
We still have 35 stars that do not exhibit {\it u}-band excess but have 
$\alpha_{IRAC}$$>$$-$2.56, suggesting that accretion processes are completed, 
but inner disks did not have time to dissipate. Finally, we compare our 
disk-locking criterion with {\it u}-band photometry. For early
fast and moderate rotators our disk-locking criterion agrees with
{\it u}-band data, with 88\% of early fast rotators and 78\% of moderate 
rotators having no {\it u}-band excess. On the other hand, {\it u}-band 
photometry is not compatible with expectations for slow rotators, 
with only 28\% of these objects showing {\it u}-band excess.   

\begin{figure}[t]  
\centering{  
\includegraphics[width=9.1cm]{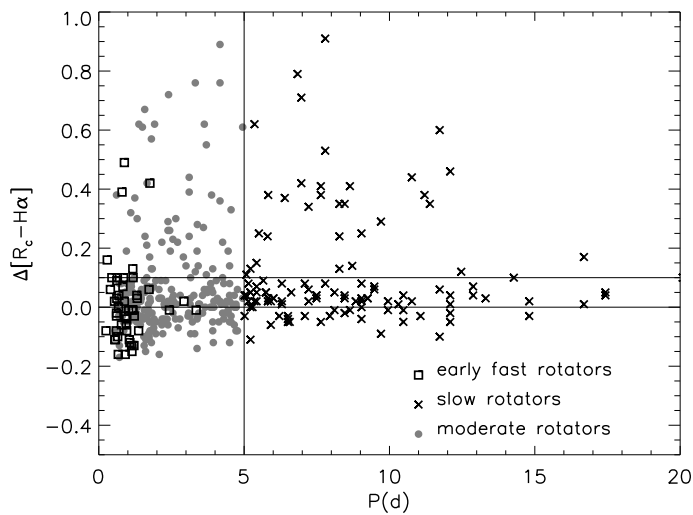}  
}  
\caption{$\Delta[R_c$$-$$H\alpha]$ versus rotation period for NGC\,2264 stars.  The symbols are the same as in Fig.~\ref{drmha}.
}  
\label{deltaper}  
\end{figure}

We now compare our disk-locking criterion 
with observational evidence for disk presence.
In this work, slow rotators (locked stars) are assumed to host circumstellar 
disks, and we expect that they are related to \textsc{ctts} and 
intermediate cases with $\alpha_{IRAC}$$>$$-2.56$. 
On the other hand, moderate rotators (unlocked objects) are assumed to be stars that have lost 
their disks in the past, and we expect that they are related to  
\textsc{wtts} and the intermediate cases that have $\alpha_{IRAC}$$\leq$$-2.56$.
\textsc{wtts} are stars without inner disks, have no accretion, 
and possibly are either objects whose disks are more evolved 
than \textsc{ctts} disks or even diskless stars.
Finally, the early fast rotators are 
stars whose disk lifetimes were shorter than $10^5$ 
years, which means that by definition, they are \textsc{wtts} with 
very short disk lifetimes and should exhibit low values of 
$H\alpha$ and $\alpha_{IRAC}$ indexes. However, they evolved practically without a disk.    
Figure\,\ref{deltaper} shows
$\Delta[R_c$$-$$H\alpha]$ versus rotation period for 
NGC\,2264 stars. The solid horizontal line at $\Delta[R_c$$-$$H\alpha]$=0.1
divides the sample into intermediate cases and classical T\,Tauri stars. The solid 
horizontal line at $\Delta[R_c$$-$$H\alpha]$=0 divides the sample into 
intermediate cases and weak-line T\,Tauri stars. 
The vertical line divides the sample into slow rotators
(P$\geq$5~days),
moderate, and fast rotators (P$<$5~days).

In Table~\ref{tabdlockcrit} we compare our disk-locking criterion with 
observational data by showing the number of stars that were classified as slow,
early fast, and moderate rotators and the number of stars of each group that were 
classified as
\textsc{ctts}, \textsc{wtts,} and intermediate cases according to their 
$H_{\alpha}$ emission. For each subgroup, we show the number of stars that 
present dust in the inner disk according to their $\alpha_{IRAC}$ index. We
also list the number of stars without $\alpha_{IRAC}$ measurements.
Of the 117 slow rotators found in our sample, only 40\% 
(\textsc{ctts} and intermediate cases with $\alpha_{IRAC}$$>$$-$2.56) have an
observational indication of a disk. However, this disagrees
with our hypothesis that stars with $P$$\geq$5~days are locked to their 
disks. Here, it is worth mentioning that one of the slow rotators
has $H_{\alpha}$ emission but no dust in the 
inner disk, which characterizes the so-called transitional disk\footnote{According
to \citet{pott10}, transitional disks are defined as systems that are
significantly depleted of optically thick dust on scales of a few astronomical
units. It is believed that the transition from optically thick accretion to
optically thin disks depleted of gas and dust is short lived \citep{hillen08}.}
class of objects.
Of the 48 early fast rotators, 87.5\% have an $H_{\alpha}$ index 
that indicates no accretion disk, as expected. In addition, $\alpha_{IRAC}$ and
$H_{\alpha}$ measurements for this group agree well.   
Of the 240 moderate rotators, 75\% have no 
indication of disks (\textsc{wtts} and intermediate cases with 
$\alpha_{IRAC}$$\leq$$-$2.56 or without $\alpha_{IRAC}$ measurements). 
Although this is what we expected, we still found 60 stars classified as 
\textsc{ctts} or
intermediate cases with $\alpha_{IRAC}$$>$$-$2.56. An alternative explanation 
for the existence of moderate rotators with observational indications of disk
presence could be that the star-disk interaction has moved inward after the 
radiative core formed \citep{gregory12}.
Thus, we can conclude that our disk-locking criterion agrees
reasonably well with observational data for early fast and moderate rotators,
but is not consistent with the data for slow rotators. 

The group of 405 objects whose periods were determined from ground-based
observations by \citet{lamm05}, and the group of 189 objects by \citet{affer13},
whose periods were determined by the CoRoT satelite, have 74 stars in common, 14 of which do 
not present the same periods in the two studies (considering the 
errors provided). By using these space-based periods, we 
again applied our criterion for a disk presence and compared the results
with the $\Delta[R_c-H\alpha]$ indexes and with $UV$ excesses by \citet{venuti14}.
This replacement of ground-based to space-based periods did not improve
the comparison between our criterion for a disk presence and the observational
disk indicators: even using CoRoT periods and more reliable disk
indicators, the two criteria for a disk presence are not consistent for slow 
rotators.
The 
115 remaining stars of the Affer sample were not included in our
study because they would prevented us from obtaining a unique observational disk 
indicator for all stars.

\begin{figure}[h]
\centering{
\includegraphics[width=8.0cm]{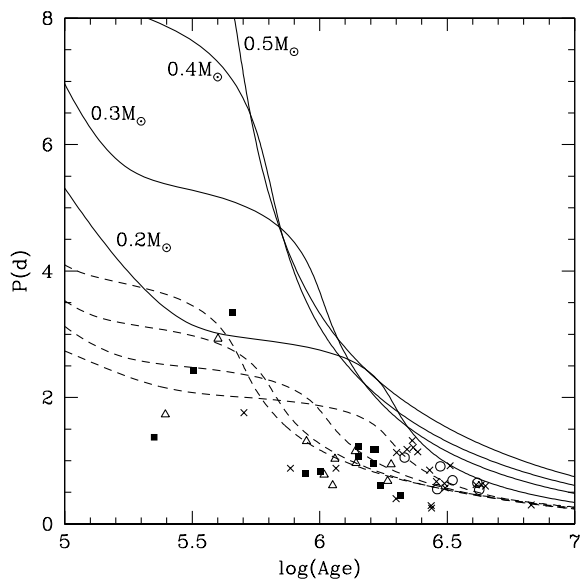}
}
\caption{Period
versus age for early fast rotator
pMS stars of NGC\,2264.
Their positions in the period-age plane
are compared with models assuming 
evolution with constant angular momentum.
Masses in the interval 0.2$<$$M$/M$_{\odot}$$<$0.3
correspond to $\circ$, 0.3$<$$M$/M$_{\odot}$$<$0.4 to $\bull$,
0.4$<$$M$/M$_{\odot}$$<$0.6 to $\triangle$
and all the remaining to $\times$.
Solid lines represent models with $J_{\rm in}$=$J_{\rm kaw}$ , while
the dashed lines stand for models with 
$J_{\rm in}$=$3J_{\rm kaw}$.
}
\label{ngcevol}
\end{figure}
  
We next analyzed the rotational evolution of NGC\,2264 
stars based on our disk-locking criterion. As 
for the ONC, we
assumed that the 48 NGC\,2264 stars
classified as early fast rotators can be considered to 
evolve with conservation of angular momentum from the
start.
This
implies
that they evolved without a circumstellar disk,
and this assumption seems to be supported by $H\alpha$ emission 
observations because 24 of these stars were classified as \textsc{wtts}, 
18 as intermediate cases, and only 6 as \textsc{ctts}.
Even using a
period as
short as $P_{\rm thresh}$, we found 48 early fast rotators
in NGC\,2264,
which
is not
very significant, statistically,
but
enough to
establish some constraints for
pMS rotational evolution.
The NGC\,2264 early fast rotators are
shown in Fig.\,\ref{ngcevol}
(similar to Fig.\,\ref{evol} for the ONC).
Again, it is necessary to assume a distribution of initial angular momenta at
least in the range $J_{\rm kaw}$$<$$J_{\rm in}$$<$3$J_{\rm kaw}$, 
models with $J_{\rm in}$=$J_{\rm kaw}$ being an upper limit for NGC\,2264 objects 
classified as early fast rotators by the criterion of hypothesis 1.
To reproduce the positions of stars with the lowest periods
in the period-age plane, it is necessary to use models with high initial angular momentum
($J_{\rm in}$$>$3$J_{\rm kaw}$), as in the work by 
\citet{barnes01}. Their solid-body models with initial periods of 4 days 
(at the birthline) and angular momentum conservation from the beginning can 
account for the slowest rotators of solar type of our sample. It is difficult to compare the initial conditions of our models and those of Yale. The starting point of the Yale models is placed at the birth line (defined on the 
deuterium main sequence), and our models start before it. For example, our 0.6\,M$_{\odot}$
calculations reach the deuterium-burning phase with a rotation period of 
about 10-15~days.  

\begin{figure*}  
\centering{  
\includegraphics[width=19.0cm]{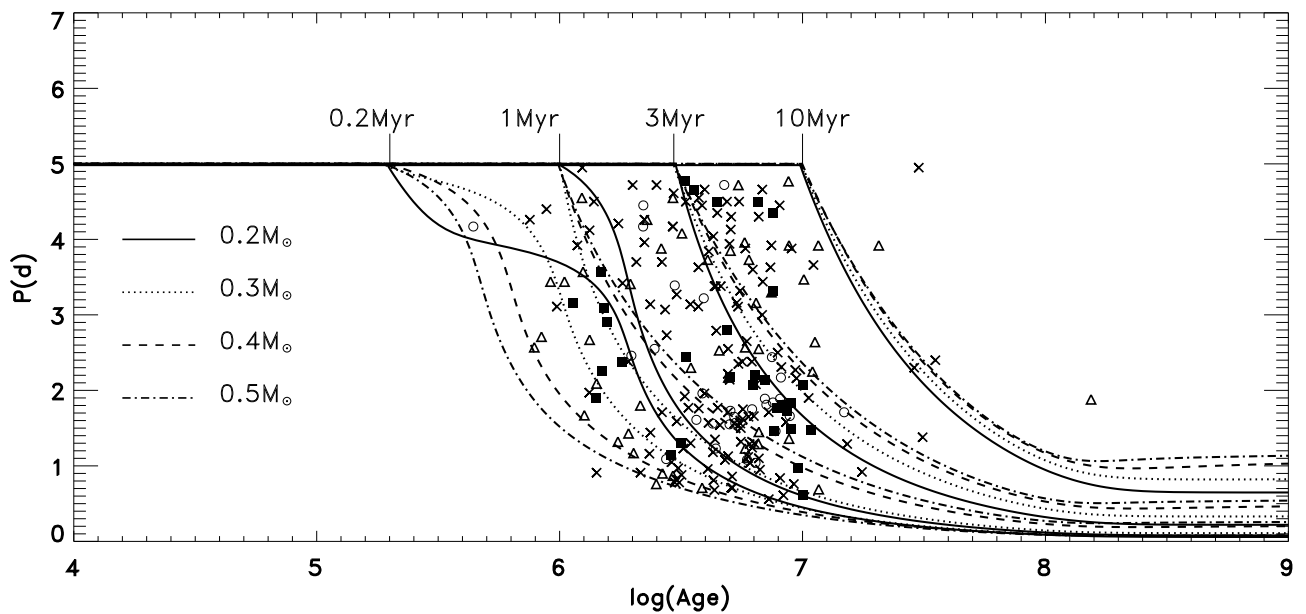}  
}  
\caption{Period  
versus age for moderate rotators  
pMS stars of NGC\,2264.  
Their positions in the period-age plane are compared with 
disk-locking models with $P_{\rm disk}$=5\,days and 0.2 (solid), 
0.3 (dotted), 0.4 (dashed), and 0.5\,M$_{\odot}$ (dash-dotted).   
The symbols are the same as in Fig.~\ref{ngcevol}.  
}  
\label{ngcevolmod}  
\end{figure*}  

We assumed that the objects classified as moderate rotators
can be used
to test the hypothesis that disk-locking is the 
mechanism preventing stellar spin up. Then, we 
used the 240
moderate rotators found in our sample, which, according
to our criterion,
are currently unlocked stars, have lost their disks in 
the past time, and for which we expect to find 
no evidence of a disk. 
Observations related to $H\alpha$ emission
show that 45 stars were classified as \textsc{ctts}, 72 as \textsc{wtts,} and
123 as intermediate cases, partially supporting this 
hypothesis. 
Although 81,25\% of the moderate rotators have 
$\Delta[R_c$$-$$H\alpha]$$<$0.1, a 
considerable number of stars (123) have low positive
$H\alpha$ indexes and 15 of these stars have dust in the inner disk, indicating
that 25\% of the moderate rotators could be \textsc{ctts}. 
  
In Fig.~\ref{ngcevolmod} we show  
NGC\,2264  
moderate rotators  
in the rotation period--age plane together with our  
models for a series of disk lifetimes, P$_{\rm lock}$=5~days  
and tracks  
for 0.2, 0.3, 0.4, and 0.5\,M$_{\odot}$. The 0.5\,M$_{\odot}$  
models were included  
because of the considerable number of  
NGC\,2264 objects in the mass range of 0.5-0.6\,M$_{\odot}$  
(see Fig.~\ref{ngchistm}),  
which are not seen in the ONC mass distribution (Fig.\,\ref{hists}).  
Figure\,\ref{ngcevolmod} shows that the observed  
positions of moderate   
rotators in the period-age plane are well reproduced by our theoretical  
models with disk-locking ($P_{\rm disk}$=5\,days and $T_{\rm disk}$=0.2,  
1, 3, and 10~Myr).   
However, disk lifetimes in the range of 5-10~Myr, which are needed to
reproduce the period distribution of a considerable number of NGC\,2264 stars, seem to be  
only poor estimates. According to \citet{bouvier07}, at 
10~Myr the fraction of stars that still  
have disk is around 10-25\%. \citet{sicilia06} reported that at an age of 5~Myr, about
90\% of disks have already dispersed, and within 10~Myr, almost all pMS stars 
are diskless. More recently, \citet{fedele10} determined 
an accretion timescale of 2.3\,Myr and a near-to-mid infrared excess timescale of 
3\,Myr, indicating that mass accretion in pMS stars seems to drop below the
detectable level earlier than near-to-mid infrared excess. 

The analysis we made in hypothesis 1, which assumes that
NGC\,2264 stars with rotation periods longer than 5~days evolved with a constant
angular velocity (locked to their disks), shows that our disk-locking 
criterion is not compatible with $H\alpha$ and $\alpha_{IRAC}$ observational 
data. In our slow rotators sample (117 objects), 30 are classified as 
\textsc{ctts} and  
69 stars are classified as intermediate cases. In the intermediate case group, 42 stars 
have $\alpha_{IRAC}$ measurements and 17 of these have $\alpha_{IRAC}$$>$-2.56 
and could be \textsc{ctts} (see Table~\ref{tabdlockcrit}). This
means that we have 
observational evidence of a disk only for 40\% of our slow rotators.  
Furthermore, it is worth remembering that according to \citet{lamm05} 
and based on the model by \citet{shu94}, estimates for the locking period 
of NGC 2264 stars indicate that a $P_{\rm lock}$=5~days
is very short for higher mass stars. This short locking period and disk lifetimes of 
5-10~Myr assumed in this section are very unlikely. As a result
of these discrepancies
in our physical hypothesis, we
investigated another scenario for the period distribution of NGC\,2264, in which \textsc{ctts} are not associated with stars with $P$$\geq$5~days. We describe this 
in the next section.

\subsection{NGC2264 stars - hypothesis 2}\label{ngc8dias}  
  
Another  
possible interpretation of the rotation evolution of NGC\,2264 stars   
was presented by \citet{lamm05}, who suggested  
that  
the  
period distribution  
of both NGC\,2264 and the ONC  
were similar  
when  
the clusters had the same age.  
In this  
view, stars in NGC\,2264 with $P$$\geq$4-5 days   
are assumed to  
have been  
locked in the past with the locking period attributed to  
the ONC ($P_\mathrm{lock}$$\sim$8~days) and released from their disks presumably when they  
were at the age of ONC stars. Since then, these stars would have spun up conserving  
angular momentum.   

To test this hypothesis, we tried to simulate the NGC\,2264 period distribution
at 1\,Myr. If we simply used the stars' estimated ages and projected them
back to 1\,Myr, we would be assuming that all the stars were formed 
together, disregarding the apparent spread on age of 1-10\,Myr seen in 
Fig.~\ref{ngchistage}. According to our age determinations, the apparent 
age dispersions in NGC\,2264 and the ONC are similar: 83\% (96\%) of ONC stars and 75\% 
(95\%) of NGC 2264 stars lie within one (two) standard deviation from their 
respective mean ages. Therefore it makes more sense to simulate the NGC 2264 period 
distribution when the mean age of its stars was 1\,Myr. 
In this way, if the current mean age of NGC\,2264 stars is 3.5~Myr,  
2.5~Myr ago the  cluster's mean age was about 1\,Myr. 
It is important to stress that this period distribution simulation of NGC\,2264 
stars at the ONC age is very simplistic because the dynamical evolution of stellar clusters
is subject to many competing physical conditions \citep[see, e.g.,][]{lada2003}; 
nevertheless, we believe it is a valid approach for obtaining some insight about the 
evolutionary history of NGC\,2264.
We then defined a quantity named    
``age difference'' to describe the difference between the stellar age (estimated in  
Sect. \ref{ngc2264} by our stellar models) and 2.5~Myr. We computed the age difference  
for all stars in our sample. However, stars in NGC 2264 are as young as 1~Myr  
and as old as 10~Myr (see Fig. \ref{hists1}); therefore, 121 stars with age difference $<$0  
had not been formed 2.5~Myr ago and they will not enter our simulation of the NGC\,2264 period  
distribution at the mean age of 1~Myr. To estimate the rotation period of NGC 2264  
stars with an age difference $\ge$0, we proceed in the following way. For the 
stars without observational evidence of disks
(\textsc{wtts} and intermediate cases with $\alpha_{IRAC}$$\leq$$-$2.56 or intermediate cases without $\alpha_{IRAC}$ measurements),
we calculated with our evolutionary tracks  
their radii at 2.5\,Myr ago and their current radii.  
By   
considering  
conservation of angular momentum and the observed rotation 
period, we also calculated their rotation period   
at 2.5~Myr ago, that is, when the mean age of NGC\,2264 stars
was
1\,Myr. For the \textsc{ctts} and intermediate cases   
with $\alpha_{IRAC}$$>$$-$2.56    
(those that probably still have  
their disks), we assumed that their rotation periods  
have not changed during  
these 2.5\,Myr, so that,  
when the mean age of NGC\,2264 was 1\,Myr, they had the same period as they have  
now.  

\begin{figure*}[ht]  
\centering{  
\includegraphics[width=8.0cm]{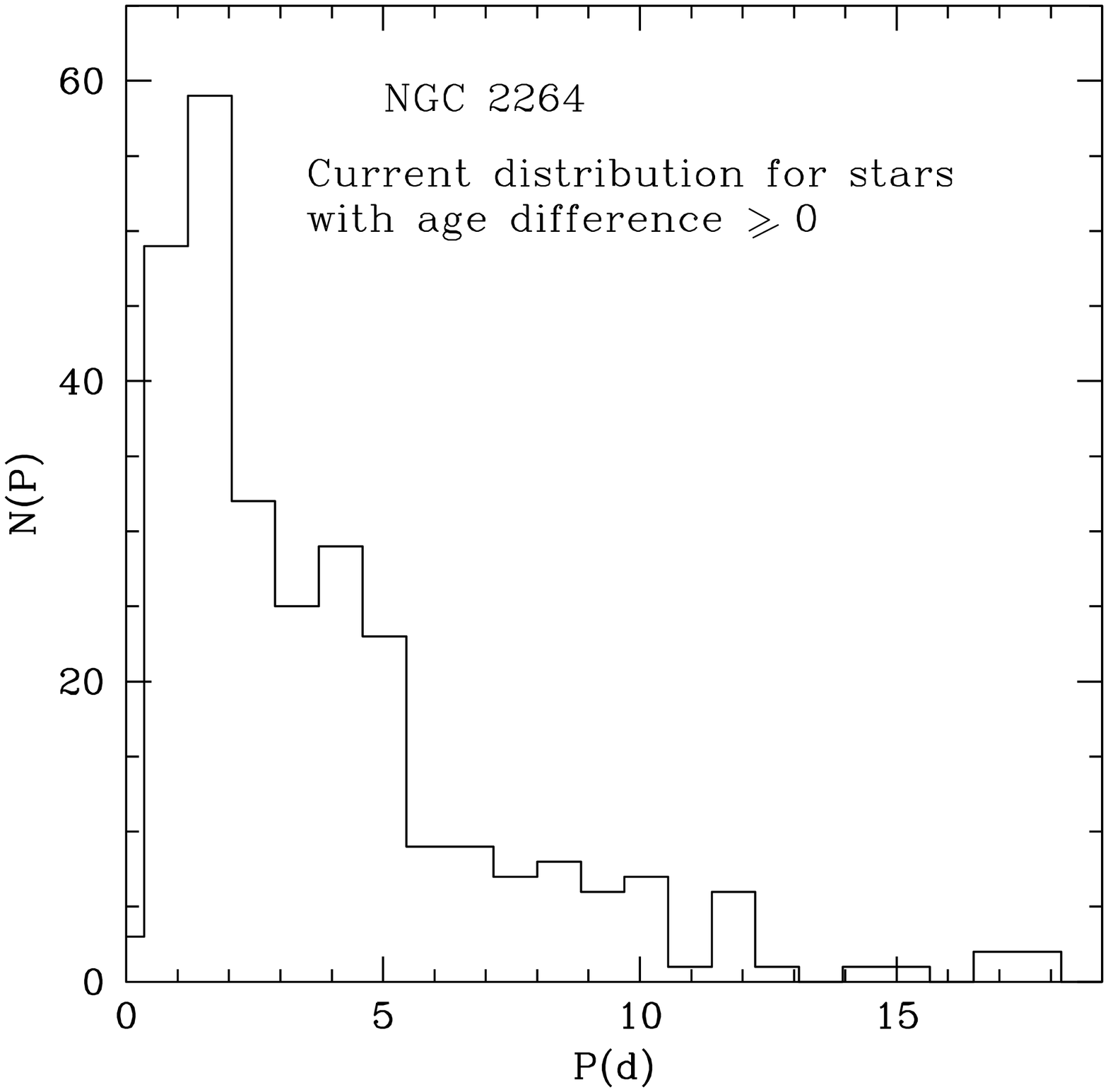}  
\includegraphics[width=8.0cm]{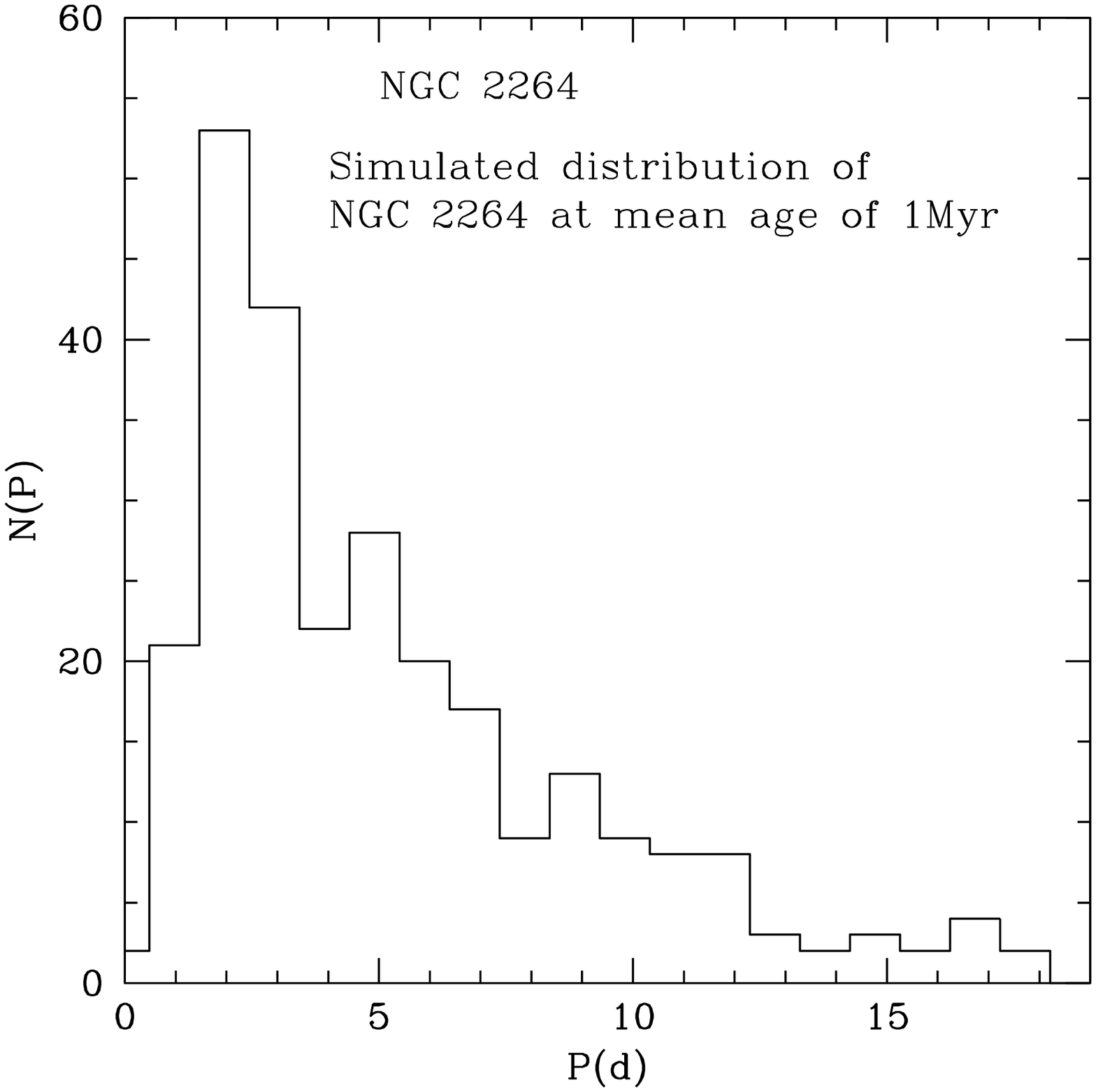}  
}  
\caption{Present period distribution of NGC\,2264 stars that probably do not have a disk  
(left panel). Simulation of the period distribution of NGC\,2264 stars when the   
objects had, on average, 1Myr (right panel).}  
\label{simuldist1}  
\end{figure*}  

The  
present and simulated period distributions  
are shown in the  
left and right  
panels of Fig.\,\ref{simuldist1},  
for comparison purposes.  
A shift in the peaks of the rotation periods between these  
two histograms  
is clearly seen.  
In the current distribution (left panel of Fig.~\ref{simuldist1}) the  
primary peak is formed by two bins in the interval of 0.5-2\,days. In the simulated   
distribution (right panel of Fig.~\ref{simuldist1}) this peak is also 
formed by two bins and is shifted to the interval of 2-3\,days.  
The secondary peak shifts from 4-5\,days to 5-7\,days.  
Our simulated period distribution for NGC\,2264 at 1~Myr is similar to that exhibited by the ONC,  
but the secondary peak of the ONC occurs around 8 days.  

\begin{figure*}[t]  
\centering{  
\includegraphics[width=8.0cm]{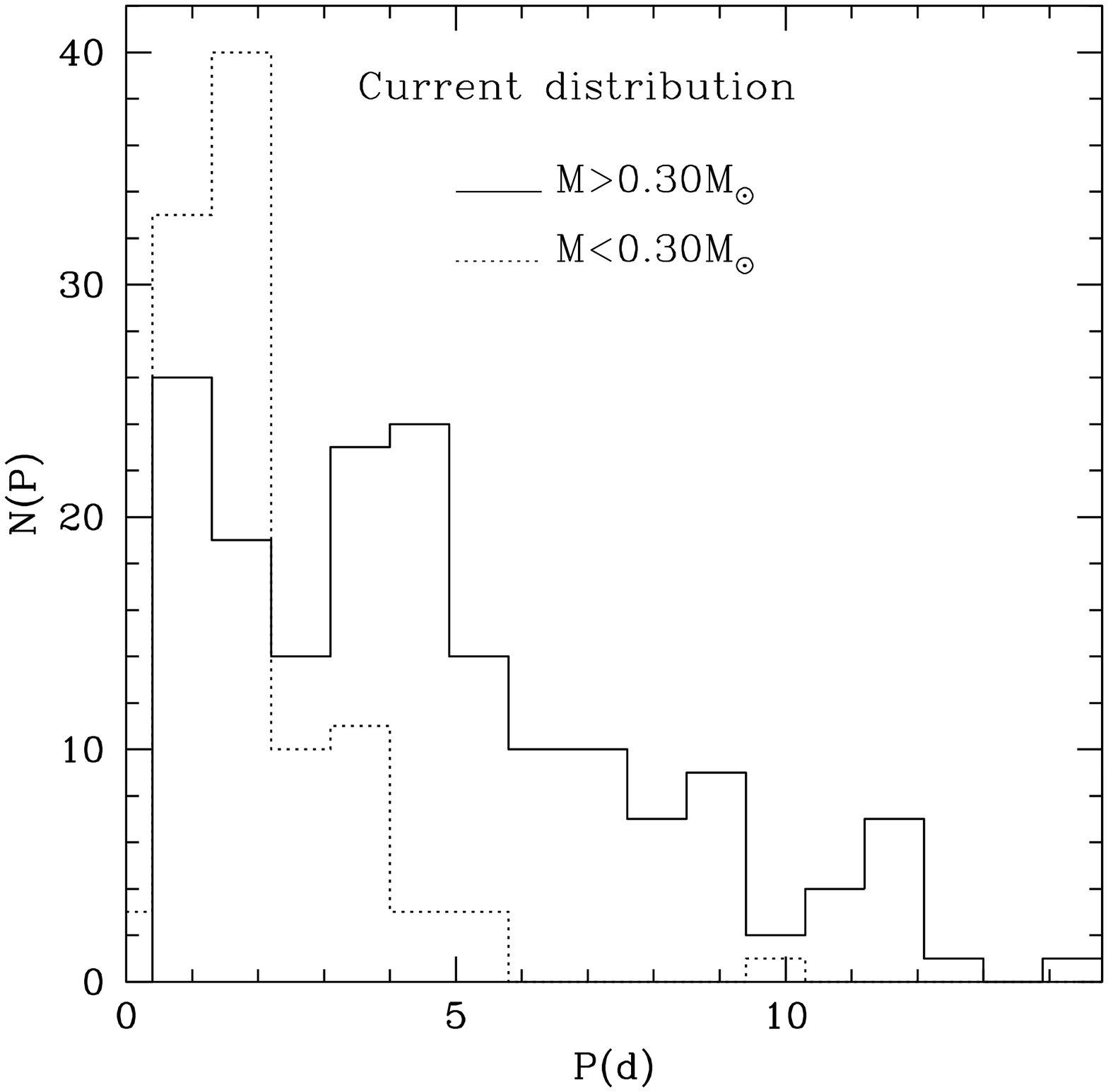}  
\includegraphics[width=8.0cm]{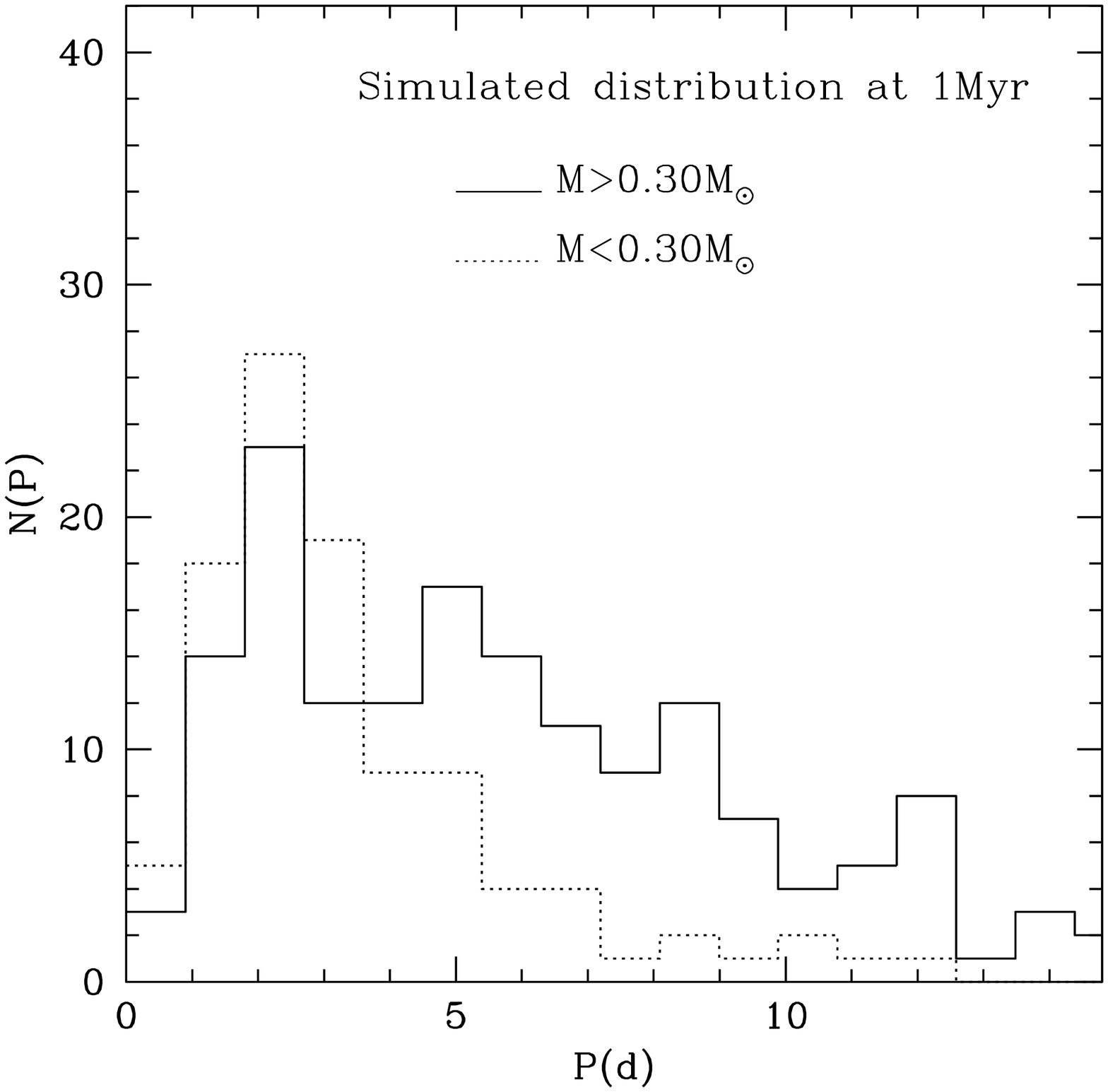}  
}  
\caption{Same as Fig. \ref{simuldist1}, but distinguishing between lower (dotted lines) and  
higher (solid lines) mass stars (M$_{\rm tr}$=0.3\,M$_{\odot}$).}  
\label{simuldist2}  
\end{figure*}

We now investigated the behavior of period distributions of NGC\,2264 stars  
by dividing the objects into the  
groups  
of low ($M$$<$0.3\,M$_{\odot}$) and   
high ($M$$>$0.3\,M$_{\odot}$) mass.  
In the left panel of Fig.\,\ref{simuldist2} we show the current period   
distributions of these stars.  
It is   
fairly similar to the right panel of Fig.\,\ref{ngcdicho},  
the difference  
is due to the absence of objects with an age difference lower than 1\,Myr  
in Fig.~\ref{simuldist2}.  
The simulated distribution (right panel of   
Fig.\,\ref{simuldist2}) agrees reasonably well with our expectation:  
a distribution with one peak around 2 days for lower mass stars, and a bimodal   
distribution for higher mass stars with one peak around 2~days and another one around  
5-7~days. The distribution shown in the right panel of Fig.~\ref{simuldist2} is   
similar to the one exhibited by the ONC. The differences remain mainly in the rotation  
period of the secondary peak, which is slightly shorter than the ONC value   
($\sim$8~days).   
  
\begin{figure*}[t]  
\centering{  
\includegraphics[width=8.0cm]{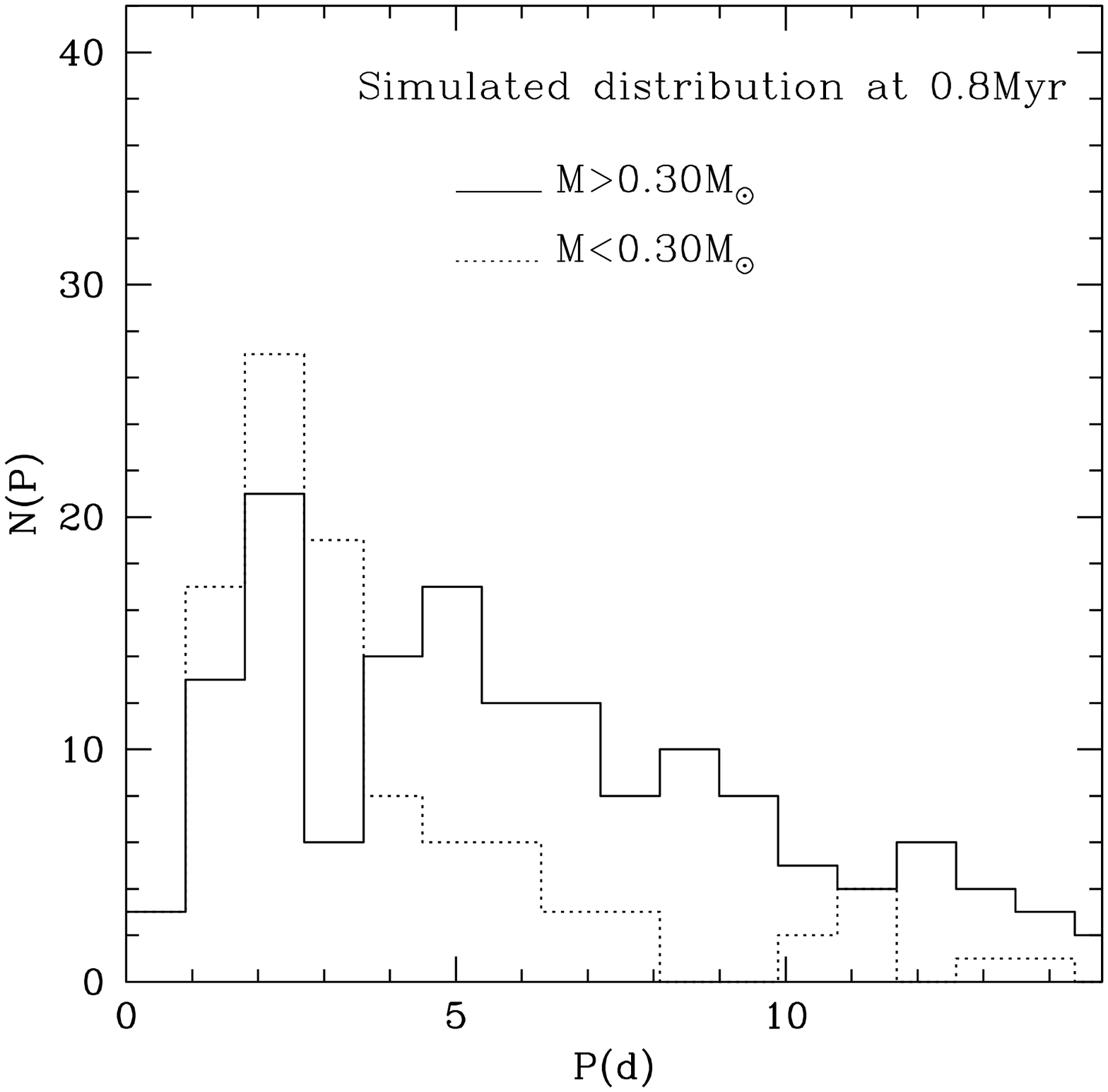}  
\includegraphics[width=8.0cm]{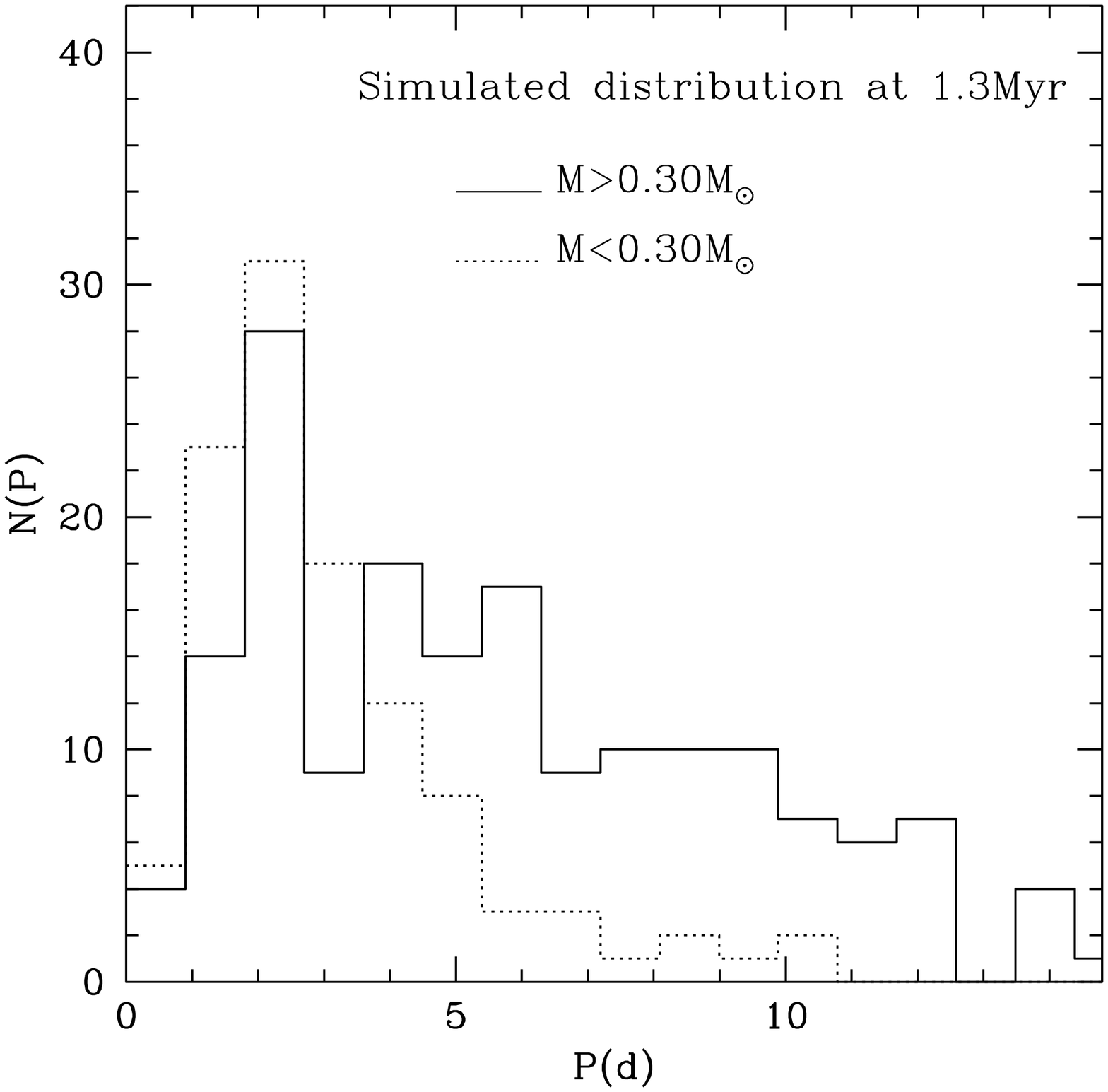}  
}  
\caption{Simulated period distributions for lower and higher mass stars   
($M_\mathrm{tr}$=0.3\,M$_{\odot}$) of NGC\,2264 at mean ages of 0.8\,Myr 
(left) and 1.3\,Myr (right).  
}  
\label{simuldist3}  
\end{figure*}  
  
Our results  
depend on the assumption that NGC\,2264 stars   
lost their disks when the mean age of the cluster was the current mean  
age of the ONC (1\,Myr).   
Studies of near-infrared excesses \citep{hillen08} for several young   
clusters  
contain information about the disk fraction of each cluster.  
Since the age of each cluster is known,  
a diagram with the fraction  
of stars with a disk as a function of time  
can be obtained and, in  
this way,  
\citet{meyer09} found that clusters with ages  
$\sim$0.8-1.3\,Myr often  
present considerable fractions of stars with disks 
(around 70\%). To take this  
into account,  
we investigated the period distribution for NGC\,2264 assuming 
its stars had lost their disks slightly before or after
the mean age of 1\,Myr. We repeated the above   
calculations and found the rotation period distribution of NGC\,2264, when  
the mean age of the cluster was 0.8~Myr and 1.3~Myr.  
These histograms are shown in the left (0.8 Myr) and right (1.3 Myr) panels 
of Fig.~\ref{simuldist3}.    
The period distribution does  
not change significantly  
for lower mass stars in either simulation. 
For the higher mass objects, the secondary peak produced for the simulated 
distribution of periods at 0.8~Myr is broader than that in the right panel of
Fig.~\ref{simuldist2}, ranging between 4-7 days.
In the simulated distribution at 1.3\,Myr, the secondary peak, formed by the
higher mass stars, ranges between 4-6 days.
Our simulated 
distributions in Figs.~\ref{simuldist2} and \ref{simuldist3}
show that
there are 89 stars (lower and higher mass stars) with periods in the range 
3.5\,days$<$P$<$8\,days in 
the simulated distribution at 0.8\,Myr, 90 stars in the simulated distribution 
at 1\,Myr, and 95 stars in the simulated distribution at 1.3\,Myr. Our simulations
do not show significant changes in the disk fraction from 0.8\,Myr to 
1\,Myr, and it increases only in 5.3\% from 1\,Myr to 1.3\,Myr. 
The secondary peak of higher mass stars at 0.8\,Myr is present,
but less evident than those obtained for the simulations at 1 and 
1.3\,Myr; several stars have periods longer than 8 days. The secondary
peak of higher mass stars of the simulated period distribution at 1.3\,Myr 
ranges from 4-6 days, which is shorter than those for 0.8 and 1\, Myr. Although
NGC\,2264 stars may have lost their disks when the age of the cluster was
in the interval of 0.8 and 1.3\,Myr, the dichotomy is clearly
highlighted by the simulated distribution
at 1\,Myr.

Even for a NGC\,2264 sample biased toward \textsc{wtts} and without 121 stars with a negative age difference, we
reproduced the period distribution of NGC\,2264 at the mean age of
1\,Myr, preserving its two main characteristics: bimodality and dichotomy
with $M_{\rm tr}$$=$$0.3\,M_{\odot}$. 
Moreover, since we obtained a simulated distribution similar to that of the ONC, 
with a secondary peak close to 8 days, our results indicate that
the hypothesis that
NGC\,2264 is in a rotational stage later
than that of the ONC as a promising explanation
for the current period distribution exhibited by NGC\,2264. In addition, this
hypothesis is consistent with the observational disk diagnostics available in
the literature.

\section{Summary and conclusions}\label{conclusions}  
We presented  
new rotating models  
that can simulate one of the effects of  
the disk-locking mechanism  
by keeping the stellar angular   
velocity constant. The action of disk-locking lasts for a given time   
($T_\mathrm{disk}$=0.2, 0.5, 1.0, 3.0, and 10.0\,Myr)  
and takes place  
in the initial stages of  
pMS evolution. 
We assumed that disk-locking is the mechanism  
that regulates the stellar   
angular velocity  
in the early stages of  
pMS  
and consequently is responsible for the very characteristic distribution of rotation  
periods of higher mass, slowly rotating T\,Tauri stars.   
Models  
considering angular momentum conservation  
were generated for  
comparison purposes and to explain the rotational evolution of fast rotators.   
We found that  
distinct rotation evolutions can alter  
not only the  
evolutionary paths followed by the tracks in the HR diagram,  
but also the mass contained in the convective envelope  
relative to the total mass and  
the rotational inertia in the convection region of the star. Disk-locking   
models produce less massive convective envelopes and lower 
rotational inertia in the convective envelope
than models with  
angular momentum conservation.   
   
We used the observed stellar populations of  
both the ONC and NGC\,2264 to test our  
pMS models. By comparing  
our tracks (obtained with our models conserving angular momentum and 
disk-locking with $T_\mathrm{disk}$=0.5 and 1.0\,Myr and $P_\mathrm{rot}$=8 and 5\,days) 
with the position of the observed objects  
in the HR diagram, we assigned  
mass and  
age to  
each star of each cluster.   
We found that the bulk of the observed   
ONC stars have masses  
and ages in the ranges  
from 0.2 to 0.4\,$M_{\odot}$  
and  
0.6~Myr$\leq$Age$\leq$2.5~Myr, for all models.  
The bulk of NGC2264 population is concentrated between 0.1 and 0.6\,$M_{\odot}$  
and between 1 and 10 Myr. According to our models, the age ratio between the two  
clusters is $\sim$3.5.  
  
We confirmed the dichotomy in the  
clusters' rotational  
properties:  
objects with $M$$<$$M_{\rm tr}$  
(whose period distribution peaks  
at short periods) and  
with $M$$>$$M_{\rm tr}$  
(with a primary peak at short periods 
and a secondary peak at longer   
periods). For disk-locking models the transition value of masses is   
$M_{\rm tr}$$\sim$0.3\,M$_{\odot}$ for both clusters.    
We also found that a high percentage ($\sim$70\%) of lower mass stars   
($M$$<$0.3\,M$_{\odot}$) rotate  
fast  
(with $P$$<$4 days for the ONC and $P$$<$2.5 days for NGC\,2264).  
  
We established a  
disk-locking criterion   
based  
on rotation periods and identified  
three groups of stars. Stars with   
$P$$\geq$$P_{\rm thresh}$  
(slow rotators) were considered to be still locked  
to the disk. Stars with   
$P$$<$$P_{\rm thresh}$ were considered unlocked, 
and we determined the time at which   
these stars have lost their disks. Some stars were found to be locked only for ages   
younger than $10^5$ years  
and were classified as early fast rotators. The remaining  
unlocked stars were identified as moderate rotators. We used a different 
threshold period  
for each cluster, $P_{\rm thresh}$= 8 days for the ONC and $P_{\rm thresh}$=5 days  
for NGC\,2264.  
We compared  
our results with an observational indicator  
for a disk that is available in the literature. Our findings for the ONC agree with  
infrared excess, and those for NGC\,2264 only  
poorly agree   
with $H\alpha$ and $\alpha_{IRAC}$ indexes. 
This might be because some of the \textsc{ctts} in NGC\,2264 
have already spun up as a result of the evolution of their magnetic field topologies. 
For NGC\,2264 \textsc{ctts}, disk indicators related to
$H\alpha$ emission do not agree well with $U$$-$$V$ and IR-excesses. 
Only 30 objects were simultaneously classified as 
\textsc{ctts} and slow rotators (representing 37\% of \textsc{ctts} and
25.6\% of slow rotators).
Moreover, the hypothesis that the higher mass stars of NGC\,2264
are locked at $P_{\rm lock}$=5~days (hypothesis 1, Sect.~\ref{ngc2264})  
implies a short  
locking period and long disk lifetimes,  
which are unlikely to occur.  
   
According to our  
disk-locking criterion,  
the lower mass stars  
probably evolved without a disk  
while the  
longer period, higher mass stars  
most likely had a disk during some time in the beginning of their evolution.  
Our results indicate that the evolution of the early fast rotators is consistent with an  
evolution conserving angular momentum for both clusters. The comparison  
of the period evolution  
predicted by the models  
with the observed values suggests that the  
 initial angular momenta  
should be at least three times higher than those provided by \citet{kawaler87}.  
On the other hand, the evolution of the moderate rotators agrees with   
an evolution with disk-locking. For ONC objects, the observed periods of moderate  
rotators can be explained by means of disk-locking models with   
0.2$\leq$$T_{\rm disk}$(Myr)$\leq$3 and $P_\mathrm{lock}$=8 days. For NGC\,2264 stars,  
although disk-locking models with 0.2$\leq T_{\rm disk}$(Myr)$\leq$10 and $P_\mathrm{ lock}$=5 days  
fit the observed periods of moderate 
rotators reasonably well, this result must be considered with 
caution. Short locking periods of 5~days and long disk lifetimes
of 10~Myr are questionable. Studies point out that at this age the fraction of stars 
with disks is considerably small ($<$10-25\%).   
  
Therefore, because hypothesis 1 for NGC\,2264 
requires odd assumptions (short locking period and long disk lifetimes), 
we tested  
hypothesis  
2 (Sect.~\ref{ngc8dias})  
proposed  
by \citet{lamm05}: when NGC\,2264 stars had the same mean age as the ONC, they  
had the same period distribution. In this scenario, NGC\,2264 stars with 
observational evidence of a disk remained locked during their
evolution, while stars with no such evidence remained locked only during the 
first 1\,Myr with $P_{\rm lock}$=8 days, and after this, they evolved  
with constant angular momentum.       
By assuming  
angular momentum conservation, we simulated the period distribution  
of NGC\,2264 stars when they had, on average, 1\,Myr. We found that the total   
distribution was shifted to longer periods, resulting in a double-peaked histogram  
similar to that observed for ONC stars, but the secondary peak of NGC\,2264 occurs for  
periods slightly  
shorter than for the ONC (around 5 to 7 days).    
In addition, we found a dichotomy in the simulated distribution at  
the same transition value, M$_{\rm tr}$=0.3\,M$_{\odot}$. 
  
Our results show that a disk-locking criterion based on
rotation periods is consistent with observations, but not for any values of
$P_{\rm lock}$ and $T_{\rm disk}$. Typical values are estimated as 
$P_{\rm lock}$$\approx$8 days and $T_{\rm disk}$$\leq$3 Myr. 
As rotation periods of pre-MS
stars seem to decrease with age, the period distribution of a young cluster
seems to evolve as it ages. In the specific case of the clusters analyzed in
this work, the period distribution of the ONC seems to represent a previous
stage of the NGC\,2264 cluster, indicating that hypothesis 2 
(Sect.~\ref{ngc8dias}) is a promising scenario to explain the
period distribution of NGC\,2264.
This interpretation is supported by Monte Carlo simulations results 
by \citet{vasconcelos15}, who used similar disk-locking hypotheses and 
reproduced the disk fraction as a function of period of ONC and NGC\,2264 stars and their accretion-rotation and mass-rotation connections.

\noindent  
\begin{acknowledgements}  
The authors thank Francesca D'Antona (INAF-OAR, Italy)  
and Italo Mazzitelli (INAF-IASF, Italy)  
for granting them full access to the \texttt{ATON} evolutionary code.  
We also thank an anonymous referee for the comments and suggestions.
The financial support from the Brazilian agencies CAPES, CNPq and   
FAPEMIG is  
gratefully acknowledged.  
\end{acknowledgements}

\end{document}